\documentclass[aps,prd,showpacs,nofootinbib,floats,floatfix,preprintnumbers,groupedaddress,twocolumn]{revtex4}

\usepackage{bm}
\usepackage{latexsym}
\usepackage{dcolumn}
\usepackage{amsmath,amsfonts,amssymb}
\usepackage{graphicx,epsfig}
\usepackage{amsmath}
\usepackage{fancyhdr}
\usepackage{hyperref}
\usepackage{graphicx,epstopdf}
\hypersetup{
	colorlinks   = true, 
	urlcolor     = blue, 
	linkcolor    = blue, 
	citecolor   = red 
}

\def\p{\partial}
\def\f{\frac}

\begin{document}
\title{Induction of chaotic fluctuations in particle dynamics in a uniformly accelerated frame}
\author{Surojit Dalui\footnote {\color{blue} suroj176121013@iitg.ac.in}}
\author{Bibhas Ranjan Majhi\footnote {\color{blue} bibhas.majhi@iitg.ac.in}}
\author{Pankaj Mishra\footnote {\color{blue} pankaj.mishra@iitg.ac.in}}

\affiliation{Department of Physics, Indian Institute of Technology Guwahati, Guwahati 781039, Assam, India
}

\date{\today}

\begin{abstract}
The ongoing conjecture that {\it the presence of horizon may induce chaos in an integrable system}, is further investigated from the perspective of a uniformly accelerated frame. Particularly, we build up a model which consists of a particle (massless and chargeless) trapped in harmonic oscillator in a uniformly accelerated frame (namely Rindler observer). Here the Rindler frame provides a Killing horizon without any intrinsic curvature to the system. This makes the present observations different from previous studies. We observe that for some particular values of parameters of the system (like acceleration, energy of the particle), the motion of the particle trapped in harmonic potential systematically goes from periodic state to the chaotic. This indicates that the existence of horizon alone, not the intrinsic curvature (i.e. the gravitational effect) in the background, is sufficient to induce the chaotic motion in the particle. We believe the present study further enlighten and balustrade the conjecture.  
\end{abstract}

\maketitle


\section{\label{Intro}Introduction and Motivation}
The intimate relationship between the geometrical properties of spacetime horizons and the dynamics of particle motion near it has always been an interest to many researchers. The horizon, which acts as one way membrane, is one of the most fascinating boundaries gravity.  These arise as the solutions of gravitational equations of motion \cite{Paddy book}. The typical examples are: Schwarzschild black hole which has {\it event horizon}, the {\it de-Sitter horizon} which emerge when the spacetime has a finite boundary at the asymptotic region, {\it etc}. These arise in the purely curved background. Interestingly, even in Minkowski spcetime, we have the {\it Rindler horizon} which is an observer (uniformly accelerated) dependent non-compact horizon.

 In last few decades, near horizon physics, at classical as well as quantum level, has been the active area of research \cite{Bombelli:1991eg,Sota:1995ms,Vieira:1996zf,  Suzuki:1996gm,Cornish:1996ri,deMoura:1999wf,Hartl:2002ig,Han:2008zzf,Takahashi:2008zh,Hashimoto:2016dfz,Li:2018wtz,Hashimoto:2018fkb}. In recent time there have been many attempts to study the particle dynamics in presence of event horizon for different kind of black hole systems where the black hole is either static spherically symmetric \cite{Sota:1995ms,Suzuki:1996gm} or rotating \cite{Hartl:2002ig} or magnetized \cite{Li:2018wtz}. The test particles considered in those cases are either massive, charged, and spinning \cite{Han:2008zzf,Takahashi:2008zh} or the massless  ones \cite{Dalui:2018qqv}. All the analysis done in those cases are performed at the classical level \cite{Bombelli:1991eg,Suzuki:1996gm,Cornish:1996ri,deMoura:1999wf,Takahashi:2008zh} and the observed results show that the motion of the test particles become chaotic in nature. There has also been an attempt to understand the behaviour of Lyapunov exponent for the case of null trajectories of a particle \cite{Cardoso:2008bp}. Another interesting observation has been made in \cite{Cornish:1996de}, where the authors studied the motion of a test particle in the field of two fixed centres described by a family of Einstein-Maxwell-Dilaton theories and showed that the transition between regular to chaotic motion is possible as the dilation coupling is varied. In the quantum regime also the influence of horizon on the particle dynamics has been studied. It has been observed that the particle's motion, detected in the Minkowski vacuum by Rindler detector, is Brownian type \cite{Adhikari:2017gyb}. A recent analysis in de-Sitter spacetime, from the perspective of quantum field theory, also demonstrated that the produced  particle seen from the comoving observer in conformal vacuum exhibits Brownian type of motion \cite{Das:2019aii}. 
  
  Moreover, at the classical level it was explored the fact that the regular motion of a system can be chaotic when it experiences the presence of horizon. This has been tested so far for a massive as well as massless particle in the backgrounds of static spherically symmetric (SSS) as well as Kerr spacetime \cite{Hashimoto:2016dfz, Dalui:2018qqv}. In particular, our previous analysis \cite{Dalui:2018qqv} has been done for the system in which a particle is trapped in two-dimensional harmonic potential for the SSS and Kerr spacetimes. We observed that the harmonic motion can get affected and that may lead to chaos in certain range of parameters of the composite system. In this regard, it must be noted that {\it all the previous analyses are done for an intrinsically curved background}, except in \cite{Hashimoto:2016dfz} where the Rindler case has been investigated for a massive particle. So in these cases, the motion of the particle got affected not only by horizon, but also by the curvature of the space-time as well as its own mass.

  On the basis of these works mentioned above, a further spontaneous and interesting question in this context that arises is: {\it what would be the role of only horizon on the overall behavior of the particle motion which lacks the presence of any intrinsic curvature in the spacetime}. So far we observed in different analysis that presence of only horizon captures important features in the system. One of the well known facts is at the quantum level the Rindler horizon is associated with temperature and entropy \cite{Unruh:1976db}, which is similar to the Hawking effect \cite{Hawking:1974rv} in black hole case. Although, the later one contains intrinsic curvature, the former does not. Therefore by taking one step further in this paper we are interested to address the unique role of horizon into the particle motion at the classical level. Other than going into the complexities of curved spacetime, one can think much more simplified geometry, where a horizon appears naturally and analyze the particle trajectories in order to get the role of horizon. So here we exploit that liberty and consider the well known Rindler spacetime which is merely a flat spacetime (as it is an uniformly accelerated frame in the Minkowski spacetime and the metric is constructed by using the suitable coordinate transformations). Our objective is to focus on the unique implications of the horizon and to see how its presence leaves an imprint into the particle motion.
In this regard, we would like to mention that the Rindler spacetime is an interesting choice as {\it these scenarios can be built up experimentally}. 
Moreover, if chaos emerges into the particle dynamics again then the previous claim -- {\it the presence of horizon is enough to make the particle motion chaotic} -- will be further balustraded.

Motivated by such interesting possibilities, here we consider a chargeless and massless particle is trapped under the influence of two-dimensional harmonic potential  in the Rindler frame. Here the harmonic trap direction is one along the Rindler $x$-direction (when the accelerated frame is moving parallel to Minkowski $X$ axis) and the other one along one of the transverse directions. The particle confined into the harmonic potential  in the uniform accelerated frame, with respect to this accelerated observer,  will also feel the effect of Rindler horizon. This type of model clearly fulfil our purpose of the present investigation as the Rindler frame does not have real curvature effect.

In this regard, it must be mentioned that the analytical calculations only for the motion of an outgoing particle along the x-direction show that position along this particular direction grows exponentially with time when it is situated very near to the horizon. This is certainly the signature that the system can experience chaos. Performing the numerical analysis for the above mentioned model and analyzing the Poincar$\acute{e}$ sections of the particle trajectories we confirm that chaos is inevitable for certain range of parameters. Another interesting result which is found is that the upper bound of the Lyapunov exponent is the acceleration of the particle which is consistent with the conjectured bound predicted by the SYK model \cite{Maldacena:2015waa}.

The paper is organized as follows. In section II, we obtain the equations of motion of our composite system. In subsection III A, these equations of motions are analyzed throughly by solving numerically and then the Poincar$\acute{e}$ sections are investigated systematically. Next in subsection III B, we present different routes to chaos arising in the presence of different strength of interaction with the horizon and the harmonic potential due to the tuning of different parameters of the  system. Subsection III C contains the  evaluation of the Lyapunov exponent numerically. In final section we conclude our work. Two appendices are also included at the end of the paper. In appendix \ref{Appendix1}, we highlight that massive particle under harmonic trap  have  route to chaos same as obtained for the massless particles. Appendix \ref{Appendix2} shows that the route to chaos for static spherically symmetric black hole is similar to our present Rindler case.

\section{\label{EOM}Rindler frame and equations of motion}
The frame of a uniformly accelerated observer is given by the Rindler metric. In ($1+3$) dimensions it is of the form:
\begin{equation}
ds^2 = -2ax~dt^2+\frac{dx^2}{2ax}+dy^2+dz^2~.
\label{1.01}
\end{equation}
The location of the Rindler horizon is $x=0$ and $a$ is the value of the acceleration. To remove the coordinate singularity at $x=0$, we apply the Painleve coordinate transformation \cite{Painleve:1921,Parikh:1999mf}
\begin{equation}
dt \rightarrow dt - \frac{\sqrt{1-2ax}}{2ax}dx~,
\label{1.02}
\end{equation}
which transforms the metric as
\begin{equation}
ds^2 = -2ax~dt^2 + 2\sqrt{1-2ax}~dtdx+dx^2+dy^2+dz^2~.
\label{1.03}
\end{equation}
The energy of a particle with respect to this frame can be determined by the inherent timelike Killing vector $\chi^a=(1,0,0,0)$ of this spacetime. It is identified as $E=-\chi^a p_a = -p_t$, where $p_a = (p_t,p_x,p_y,p_z)$ is the four momentum of the particle.

Below, we want to mention the reasons for applying Painleve coordinate transformation.
	\begin{itemize}
		\item In our manuscript, the motive is to study the dynamics of a massless and chargeless particle in a uniformly accelerated frame and the effect of the Killing horizon on its motion. In order to do so, we confined the particle very near to the horizon so that the effect of horizon is maximum. Accordingly all the calculation will be done in the near horizon approximation. In this scenario, the form of the metric (\ref{1.01}) which has a coordinate singularity at $x=0$ is not a good choice. So one needs to choose a different coordinate system where such coordinate singularity does not appear. Also one must be careful that the non-singular coordinates should be such that the corresponding observer can see the horizon as one way membrane. This is because in our case the motion of the test particle should be affected by it. Therefore, although Kruskal coordinates system does not have the singularity problem, the corresponding observer can not feel the horizon. So such system is ruled out. On the other hand Painleve coordinate system has no such problem. It is free of coordinate singularity as well as the observer can see the horizon as one way membrane. This is also the reason for using the Painleve coordinates in discussing Hawking effect by tunneling mechanism in the null geodesic approach (see Ref. \cite{Parikh:1999mf}).
		
		\item Another important feature of Painleve coordinate transformation is that the component form of the timelike Killing vector which defines the horizon as well as energy of the particle, remains identical to that in original Schwarzschild like coordinates.
		Initially, the metric (\ref{1.01}) has the timelike Killing vector $K^{a}=(1,0,0,0)$ and the component form of this remains same in Painleve also: $\chi^{a}=(1,0,0,0)$. This can be checked by using the tonsorial transformation of a vector under coordinate transformations. This helps us to define energy in terms of only time component of four-momentum in both the coordinate systems. For instance in Schwarzschild like coordinates it is given by  $E=-K^{a}p_{a}=-p_{t}$ where $t$ is Schwarzschild time coordinate; while this is again given by only time component of momentum: $E=-\chi^{a}p_{a}=-p_{t}$ where $t$ is the Painleve time. Therefore calculations do not get complecated. 
\end{itemize}

To find the expression for energy $E$ in terms of the space components of momentum, we shall use the standard dispersion relation $g^{ab}p_ap_b=-m^2$ where $m$ is the mass of the particle. Under the background metric (\ref{1.03}), it is expanded as
\begin{equation}
E^2+2\sqrt{1-2ax}~p_xE-\Big(2ax~p_x^2+p_y^2\Big)=m^2~,
\label{1.04}
\end{equation}
where, we have considered only the motion of the particle along $x$ and $y$ directions. This particular choice is taken for simplicity by keeping in mind the minimum dimensions required to demonstrate our particular goal.
The solution of the above equation will lead to the expression for energy, which turns out to be
\begin{equation}
E=-\sqrt{1-2ax}~p_x\pm\sqrt{p_x^2+p_y^2+m^2}~.
\label{1.05}
\end{equation}
In the above, positive sign refers to the outgoing trajectories, while negative sign signifies ingoing trajectories. Since ingoing particles are trapped inside the horizon and on the other hand we are interested to investigate the system when it is very near, but outside the horizon, we shall concentrate only on the outgoing ones (positive sign  of the above equation).

Before proceeding to the main goal, we want to investigate the behaviour of particle trajectories qualitatively when it is under influence of only the Rindler horizon. For simplicity, we consider the particle to be massless; i.e. $m=0$. If one concentrates only on the motion along $x$ direction (i.e. for constant values of $y$ and $z$), then the rate of $x$ variation is given by
\begin{equation}
\dot{x}=\frac{\partial E}{\partial p_x}= -\sqrt{1-2ax}+1~.
\label{1.06}
\end{equation}
Now if the particle is very near to the horizon such that $2ax<1$, then the above Eq.(\ref{1.06}) reduces to
\begin{equation}
\dot{x}\simeq ax~.
\label{1.07}
\end{equation}
In the above Eqs.~(\ref{1.06}-\ref{1.07}), the parameter with respect to which the derivative has been defined is some affine parameter, say $\lambda$. It should be such that the four momentum of the massless particle is defined by $p^a = dx^a/d\lambda$ which satisfies $p^a\nabla_a p^b=0$. This is the usual prescription to define the parameter in the case of a null-like path $x^a=x^a(\lambda)$. Remember that $\lambda$ is similar to proper time for a timelike geodesic. The solution of the above equation is 
\begin{equation}
x=\frac{1}{a}e^{a\lambda}~.
\label{1.08}
\end{equation}
This solution implies the exponential growth of position coordinate along the $x$ direction which signifies possible induction of  chaos in an integrable system when it comes under the influence of Rindler horizon. We shall show explicitly in the later part of the paper that if this massless particle is kept in a harmonic potential in the accelerated frame, the motion of the particle in the composite system becomes chaotic for certain values of the parameters, like  total energy of the system $(E)$ and the acceleration $(a)$ of the frame. 

In order to fulfil our claim now let us concentrate on the equations of motion and see what happens when the influence of horizon comes into play. For that we consider two dimensional harmonic potential in accelerated frame which have the nature  $(1/2)K_{x}(x-x_{c})^{2}$ and $(1/2)K_{y}(y-y_{c})^{2}$ along Rindler $x$ and $y$ directions, respectively. Here $K_{x}$ and $K_{y}$ are the spring constants while $x_{c}$ and $y_{c}$ are the equilibrium positions of these two harmonic potentials.  Our goal is to discover the fact about the collective impact of horizon on this integrable system. In particular we want to inspect thoroughly in order to determine how the nature of the particle trajectory gets changed or influenced by the presence of Rindler horizon. 

We begin our calculation by obtaining the total energy of the composite system when our massless test particle moves in the background Eq.~(\ref{1.03}) under the influence of these two harmonic potentials. So the total energy of the system turns out to be (considering only the outgoing paths)
\begin{eqnarray}
E=&&-\sqrt{1-2ax}~p_x +\sqrt{p_x^2+p_y^2}+\f{1}{2}K_{x}(x-x_{c})^{2}\nonumber\\
&&+\f{1}{2}K_{y}(y-y_{c})^{2}~,\label{1.09}
\end{eqnarray}       
and correspondingly, the equations of motion will be
\begin{eqnarray}
&&\dot{x}=\f{\p E}{\p p_{x}}=-\sqrt{1-2ax}+\f{p_{x}}{\sqrt{p_{x}^{2}+p_{y}^{2}}}~;\label{1.10}\\
&&\dot{p_{x}}=-\f{\p E}{\p x}=-\f{a}{\sqrt{1-2ax}}p_{x}-K_{x}(x-x_{c})~;\label{1.11}\\
&&\dot{y}=\f{\p E}{\p p_{y}}=\f{p_{y}}{\sqrt{p_{x}^{2}+p_{y}^{2}}}~;\label{1.12}\\
&&\dot{p_{y}}=-\f{\p E}{\p y}=-K_{y}(y-y_{c})~.\label{1.13}
\end{eqnarray}  
In the next section of the paper we will examine these equations with the help of numerical analysis in order to study the motion of the particle. Here we would like to mention  that the interaction between the horizon and the harmonic potentials are taken to be so small that one can ignore that term compared to the other terms. Hence the total energy of the composite system, in the above,  has been taken as the sum of the gravitational and harmonic potentials parts.
\section{\label{NA}Numerical analysis}
To illustrate that the presence of horizon is the origin of chaos, this section will be dedicated to analyze the particle motion numerically. In the previous section we analytically showed that the particle's position along the $x$ direction grows exponentially as the particle approaches  near to the horizon. This particular feature, we invoked, as the possible induction of chaos in an integrable system.  In order to confirm our claim we will  analyze the Poincar$\acute{e}$ sections and the power spectral densities of our model (harmonic oscillator in an uniformly accelerated frame), obtained by solving the dynamical equations of motion (Eqs. (\ref{1.10})-(\ref{1.13})).  This will give us the glimpse about the change in the dynamics of the particle trajectory.  First we will present the Poincar$\acute{e}$ sections of our composite system. In the upcoming subsections we will present power spectral density and Lyapunov exponent and show how our dynamical system becomes chaotic for some particular values of parameters, like acceleration $(a)$ and energy $(E)$.

\subsection{\label{PS}Poincare sections}
\begin{figure}[!ht]
 \centering
 \includegraphics[scale=0.18, angle=0]{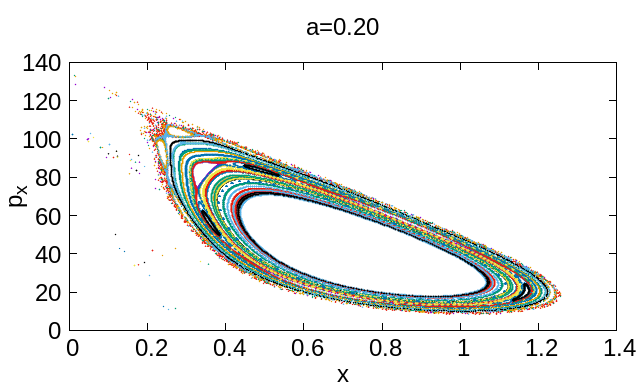}
 \includegraphics[scale=0.18, angle=0]{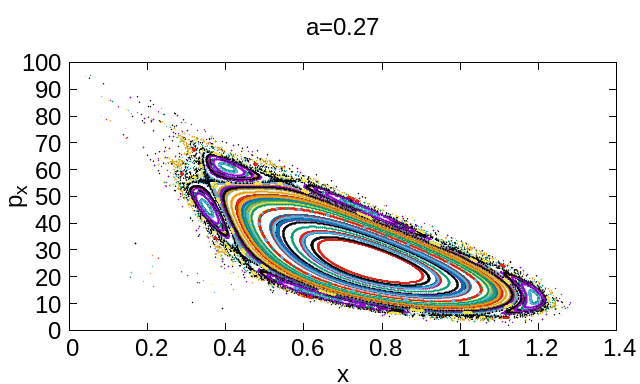}
 \includegraphics[scale=0.18, angle=0]{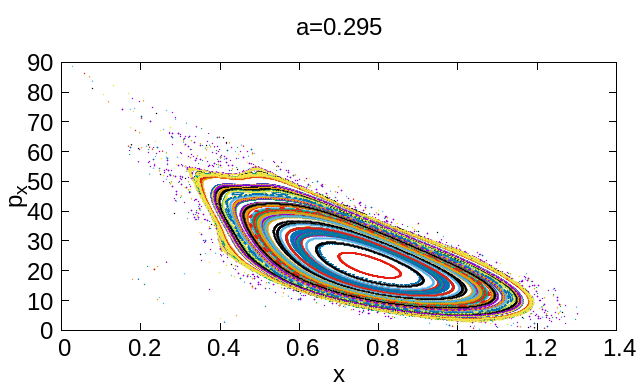}
 \includegraphics[scale=0.18, angle=0]{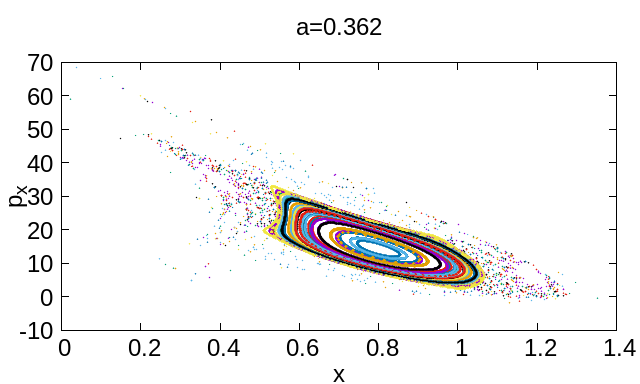}
 \caption{(Color online) The Poincar$\acute{e}$ sections in the $(x,p_{x})$ plane with $y=1.0$ and $p_{y}>0$ at different values of acceleration of the system for fixed energy ($E=24.0$). The values of accelerations are $a=0.20, 0.27, 0.295$ and $0.362$. The other parameters are $K_{x}=26.75,\, K_{y}=26.75,\,x_{c}=1.1$ and $y_{c}=1.0$. For large value of acceleration the KAM Tori break and the scattered points emerge which indicates the onset of chaotic dynamics.}
 \label{fig:Poinc_E24}
\end{figure}

\begin{figure}[!ht]
 \centering
 \includegraphics[scale=0.18, angle=0]{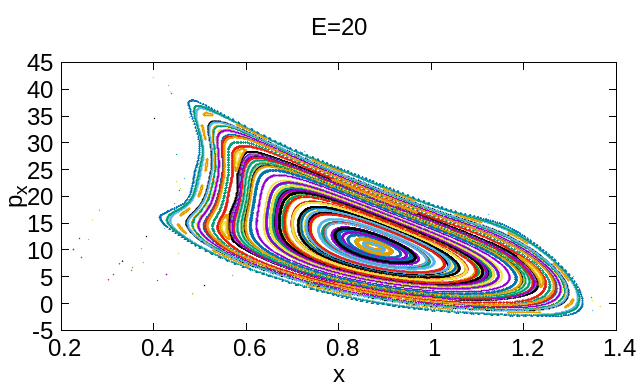}
 \includegraphics[scale=0.18, angle=0]{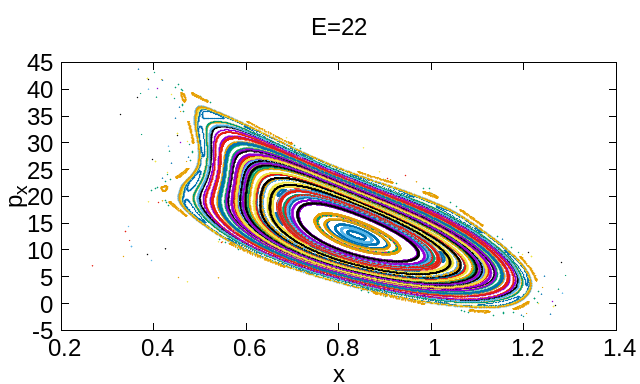}
 \includegraphics[scale=0.18, angle=0]{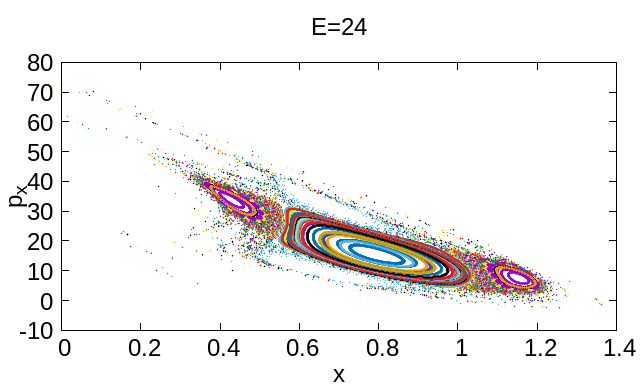}
 \includegraphics[scale=0.18, angle=0]{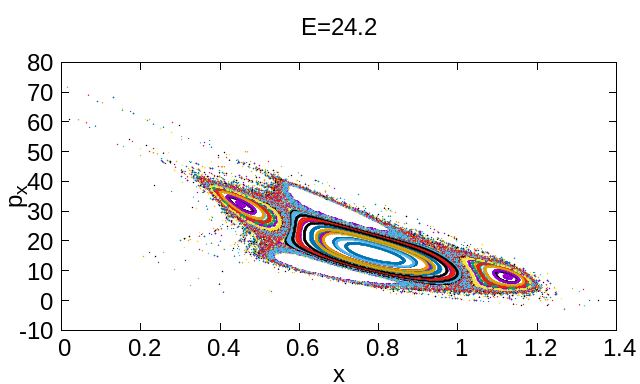}
  \caption{(Color online) The Poincar$\acute{e}$ sections in the $(x,p_{x})$ plane with $y=1.0$ and $p_{y}>0$ at different values of energy of the system but for fixed acceleration ($a=0.35$). The energies are $E=20,22,24$ and $24.2$. The other parameters are $K_{x}=26.75,\, K_{y}=26.75,\,x_{c}=1.1$ and $y_{c}=1.0$. For large value of energy the KAM Tori break and the regions filled with scattered points which indicates the presence of chaotic motion in the particle dynamics. }
 \label{fig:Poinc_a0p35}
\end{figure}

The results for the Poincar$\acute{e}$ sections of the particle trajectory are shown in Fig. \ref{fig:Poinc_E24} and Fig. \ref{fig:Poinc_a0p35}. In Fig. \ref{fig:Poinc_E24} the Poincar$\acute{e}$ sections are projected over the ($x,p_{x}$) plane for different values of accelerations but for a particular value of energy $(E)$. Whereas, in Fig. \ref{fig:Poinc_a0p35} the Poincar$\acute{e}$ sections are plotted for different values of energies $(E)$ while the acceleration $(a)$ of the system remains constant.

The dynamical equations are solved (Eqs. (\ref{1.10})-(\ref{1.13})) numerically using the fourth order Runge-Kutta method with fixed $dt=5\times10^{-3}$. The other parameters we have chosen are $K_{x}=26.75$, $K_{y}=26.75$, $x_{c}=1.1$ and $y_{c}=1.0$. The variables $x$, $y$ and $p_{x}$ are initialized with the random numbers and $p_{y}$ is obtained from Eq. (\ref{1.09}) for a fixed values of $E$ and $a$. The sections are the slice defined by $y=1.0$ and $p_{y}>0$. In Fig. \ref{fig:Poinc_E24}, the Poincare sections are plotted for different values of $a=0.20, 0.27, 0.295$ and $0.35$ while the energy is fixed at $E=24.0$ as indicated in the plots. In Fig. \ref{fig:Poinc_a0p35}, we have considered the energies $E=20, 22, 24$ and $24.2$ while the value of $a$ is fixed at $0.35$.  For different initial conditions the different trajectories of the particle are indicated by different colors in the figures.

Now let us investigate these two sets of figures in more careful way. First we explore the effect of acceleration on the particle dynamics. For low value of acceleration $a=0.20$ the display of regular KAM (Kolmogorov-Arnold-Moser) tori \cite{nonlinear:02} (see Fig. \ref{fig:Poinc_E24}) reveals the fact that the particle trajectory is yet to come under the influence of the horizon. As we increase the value of $a$ the trajectories start getting deformed and breaking into small tori. Note that the effect is more pronounced for the orbits which are closer to the horizon. There always have been some regions of regular tori in the figures instead of the fully chaotic region where all the points should be scattered. However, in the present study we could not get the fully developed chaos as we find that the orbits are quite sensitive to the parameters i.e $a$ and $E$. On further change in those parameters lead to the unbounded situations.

It is noted here that as we increase $a$, more and more regular tori are broken and correspondingly the chaotic nature of the system appears. This can be attributed to the effects of horizon on the orbit of the particle. The reason can be described in a qualitative way. The relation between the Minkowski coordinates $(T,X)$ and Rindler coordinates $(t,x)$, for the uniformly accelerated observer on the right wedge, are
\begin{equation}
a T= \sqrt{2ax}\sinh(at); \,\,\,\ a X=\sqrt{2ax}\cosh(at)~.
\label{R2}
\end{equation}
So the trajectory of the accelerated observer is
\begin{equation}
X^2-T^2 = \frac{2x}{a}~,
\label{R3}
\end{equation}
which is hyperbolic in nature for a constant $x$.
Now since $a$ is the proper acceleration (as measured by the Rindler frame), the path of the Rindler observer in terms of its own coordinate  is $(2x)/a=1/a^2$; i.e. $x=1/(2a)=$ constant. So different value of $a$ corresponds to different accelerated observer. It shows that if we increase the value of $a$, the frame will be more close to the horizon $x=0$ and the particle  will get more effect of the horizon. Hence the instability in the motion of the particle along $x$ will increase according to (\ref{1.08}). This causes breaking of the orbit when it approaches towards the horizon. That's why we here obtained more breaking of tori as we increase the acceleration.

We obtain from Fig. \ref{fig:Poinc_a0p35} that for low energy $E=20$ the Poincar$\acute{e}$ section exhibits the regular KAM  tori \cite{nonlinear:02}. As the total energy of the system gets increased the volume of the phase space increases and the trajectory of the particle approaches nearer to the Rindler horizon. As soon as the particle moves nearer to the horizon the trajectories start getting distorted into their structure which can be clearly seen from the Fig. \ref{fig:Poinc_a0p35}. The appearance of the scattered points in the outer most regions as we increase the total energy of the system indicates the presence of chaos into the system. The value of energy is chosen in such a way that the particle come close enough to the horizon but not fall into it. Finally we mention that if the particle is massive, the situation does not change. This has been confirmed numerically in Appendix \ref{Appendix1}.

          
\subsection{\label{PSD}Power Spectral Density}
In order to have an instinctual idea about the dynamical behavior of our composite system in an accelerated frame and to understand the collective impact of horizon in the particle trajectories, we systematically investigate the evolution of the position coordinate, $x(\lambda)$ with respect to the affine parameter (which plays the role of time for the massless particle). To get an extensive idea of the trajectory of the system and the deeper understanding of the onset chaotic behavior of the system, the power spectral density (PSD) is analyzed which is defined as \cite{Stoica},
\begin{eqnarray}
\text{PSD}=\f{1}{2\pi\mathcal{N}}|x(\mathcal{N},f,\Delta\lambda)|^{2}~,\label{1.14}
\end{eqnarray}   
where $x(\mathcal{N},f,\Delta\lambda)$ is the discrete Fourier transform of $x(\lambda)$ evaluated at $\lambda=k\,\Delta \lambda~(k=0,1,....,\mathcal{N}$ and $\mathcal{N}$ is the length of the discrete affine parameter series).


\begin{figure}[!ht]
	\centering
	\includegraphics[scale=0.22, angle=0]{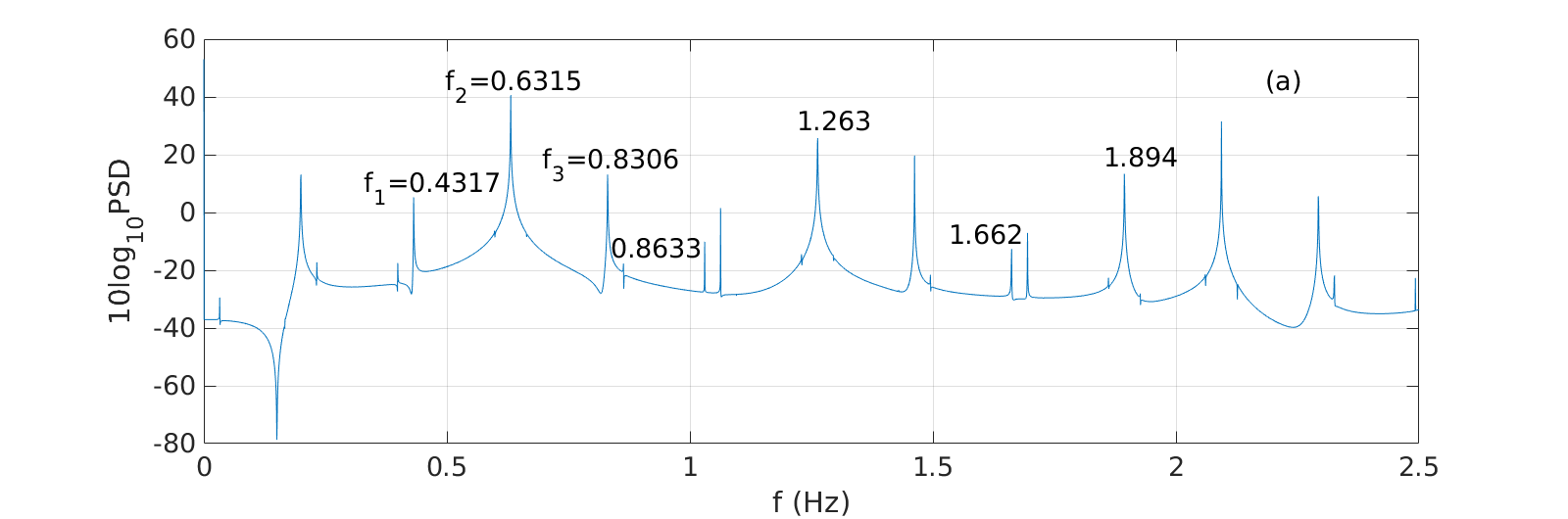}
	\includegraphics[scale=0.22, angle=0]{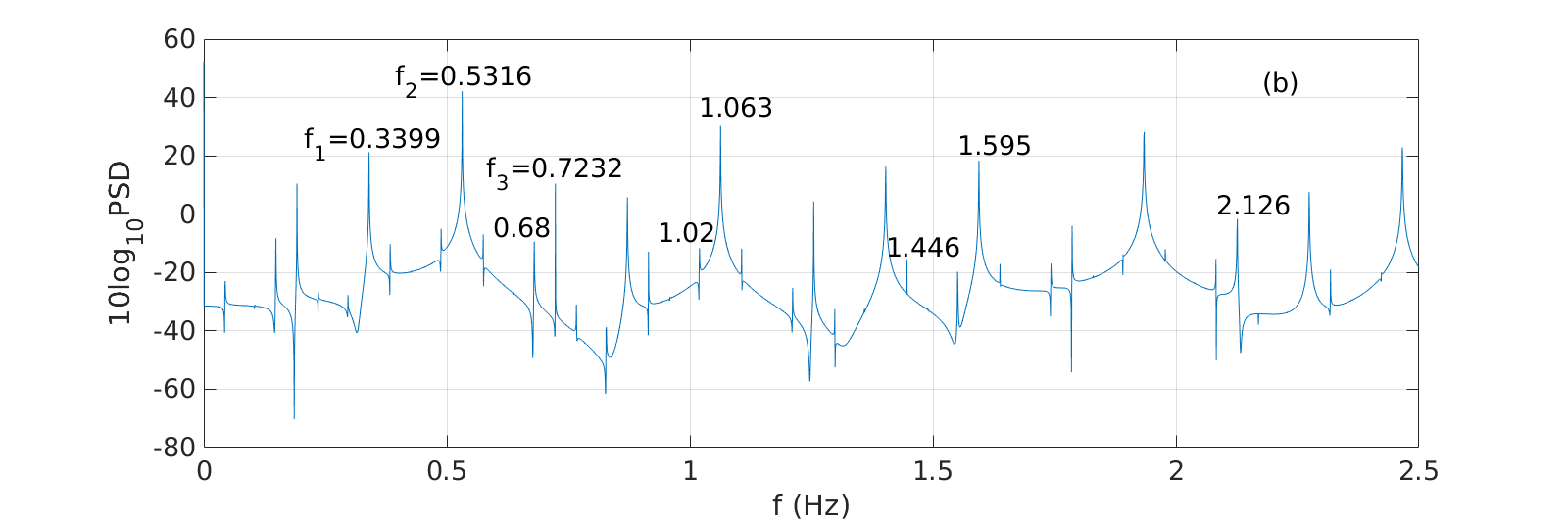}
	\includegraphics[scale=0.22, angle=0]{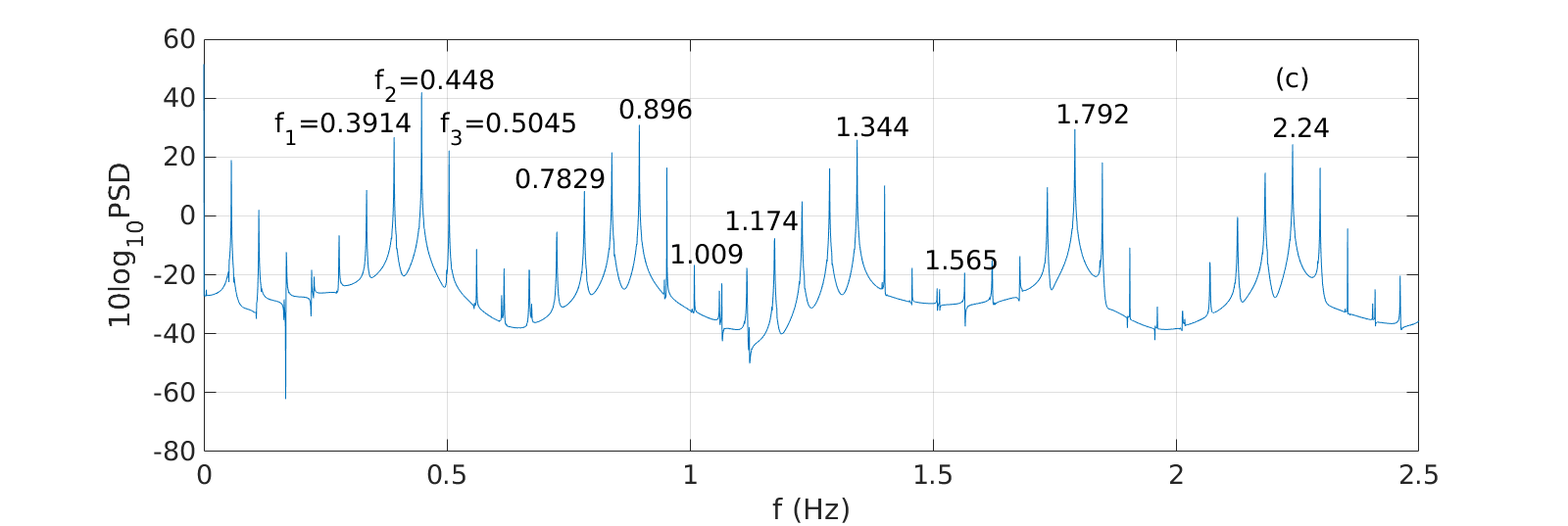}
	\includegraphics[scale=0.22, angle=0]{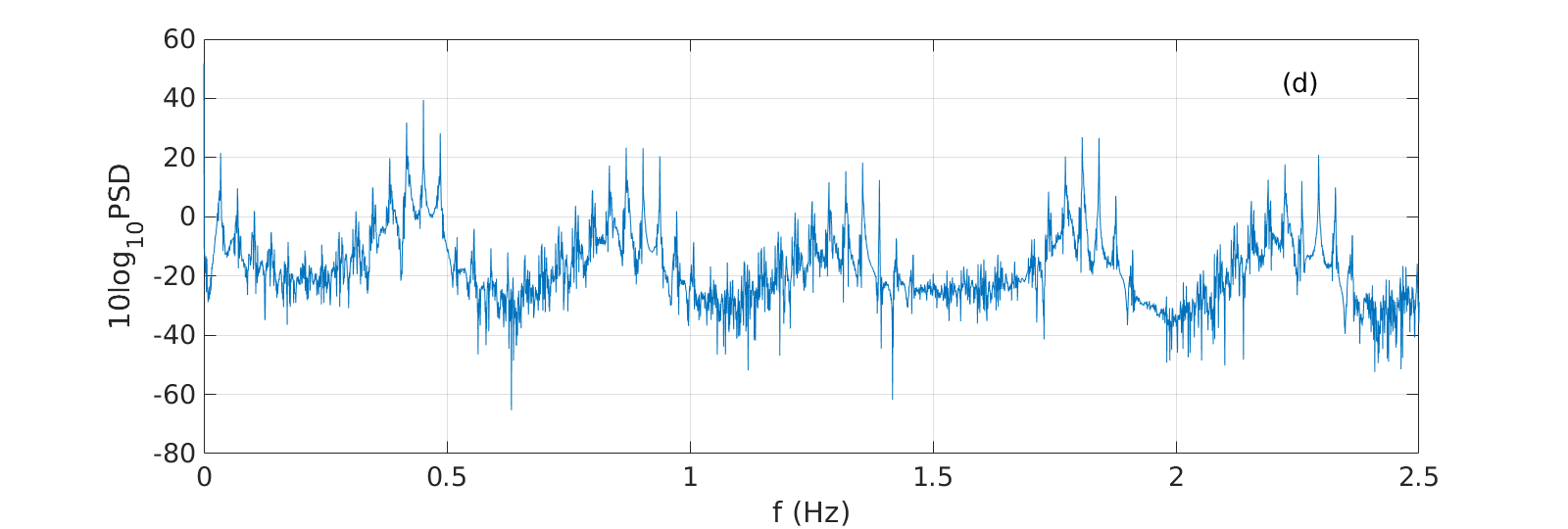}
	\includegraphics[scale=0.22, angle=0]{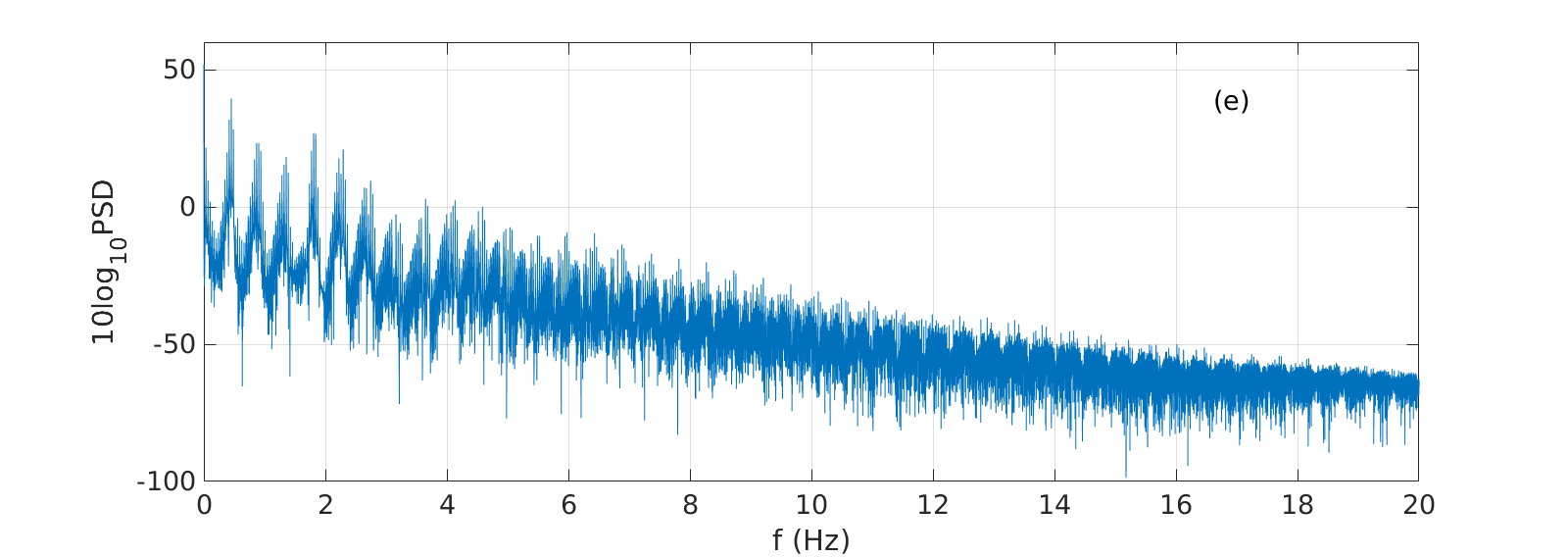}
	\caption{(Color online)  PSD for $a=0.35,\,y=1.0,\, p_{y}>0,\,K_{x}=26.75,\, K_{y}=26.75,\,x_{c}=1.1,\,y_{c}=1.0$ and for different values of the total energy of the  system $E$, namely $(a)\,E=20,\,(b)\,E=22,\,(c)\,E=24,\,(d)\,E=24.2$. Upto $E=22$, only $f_{1},f_{2}$ and $f_{3}$ and their harmonics are present but for higher values of energy i.e from $E=24$ onwards more frequencies start populating the spectrum. At $E=24.2$, the highly population of frequency spectrum which indicates the onset of chaos (see Fig. $\ref{fig:PSD_a0p35}(e)$).}
	\label{fig:PSD_a0p35}
\end{figure}

Let us see what happens when the total energy of the system $(E)$ is changed. In Fig. \ref{fig:PSD_a0p35} we show PSD for the case when the acceleration is fixed $(a=0.35)$ and $E$ is varied $(E=20, 22, 24\, \text{and}\, 24.2)$. As we found from the previous section that for lower value of energy $E=20$ the Poincar$\acute{e}$ sections are periodic in nature, that means the effect of horizon has not appeared into system yet. This feature is seen from Fig. \ref{fig:PSD_a0p35}$(a)$ where the PSD diagram is plotted for energy $E=20$ but only a few number of peaks appear into the plot. The appearance of PSD peak at $f_{2}\simeq0.6315$ is the fundamental frequency and other frequencies at $1.263, 1.894~\text{and}~2.493$ are the harmonics of that fundamental frequency. The appearance of other frequencies i.e $f_{1},f_{3}$ etc, indicates the involvement of other frequencies of oscillations in the time evolution of $x(\lambda)$, which is a signature of the quasi-periodic nature of the system. With further increase in the value of $E$, we find that more frequencies near $f_{1},f_{2}~\text{and}~f_{3}$ start getting populated (see Fig. \ref{fig:PSD_a0p35}(b) and Fig. \ref{fig:PSD_a0p35}$(c)$). Finally for high value of energy, i.e., $E=24.2$, PSD shows fully populated pattern over the frequency range (see Fig. \ref{fig:PSD_a0p35}$(d)$). That is the distinguishing character of the onset of chaos. Interestingly this has an exponential decaying character. The similar trend of PSD was also obtained in the case of the chaotic region of fluid motion \cite{Pankaj1}. In this case also, like ours, the chaos is temporal one. It may possible that such exponential decay of PSD (which is completely a phenomenological feature) is a very general nature for a system when it is in chaotic regime. But it needs further investigation which is beyond the scope this present paper. Thus for $E=24.2$, the observed a periodic time evolution of $x(\lambda)$ displays chaotic behaviour for the particle dynamics in presence of horizon.

\begin{figure}[!ht]
	\centering
	\includegraphics[scale=0.22, angle=0]{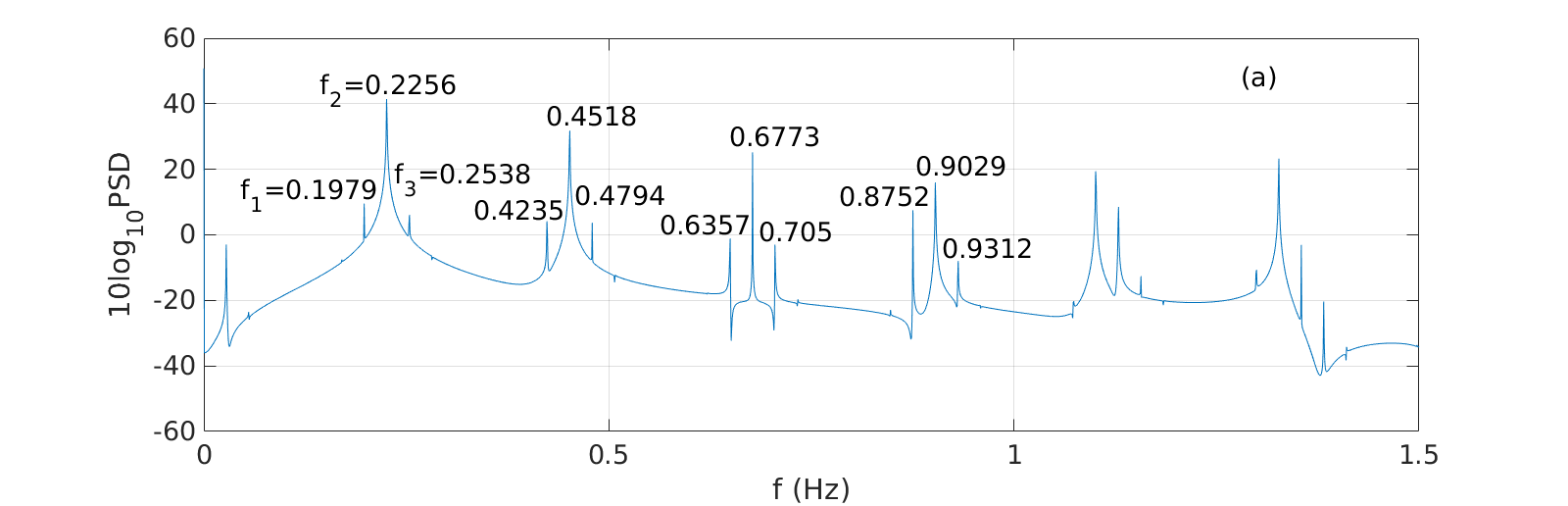}
	\includegraphics[scale=0.22, angle=0]{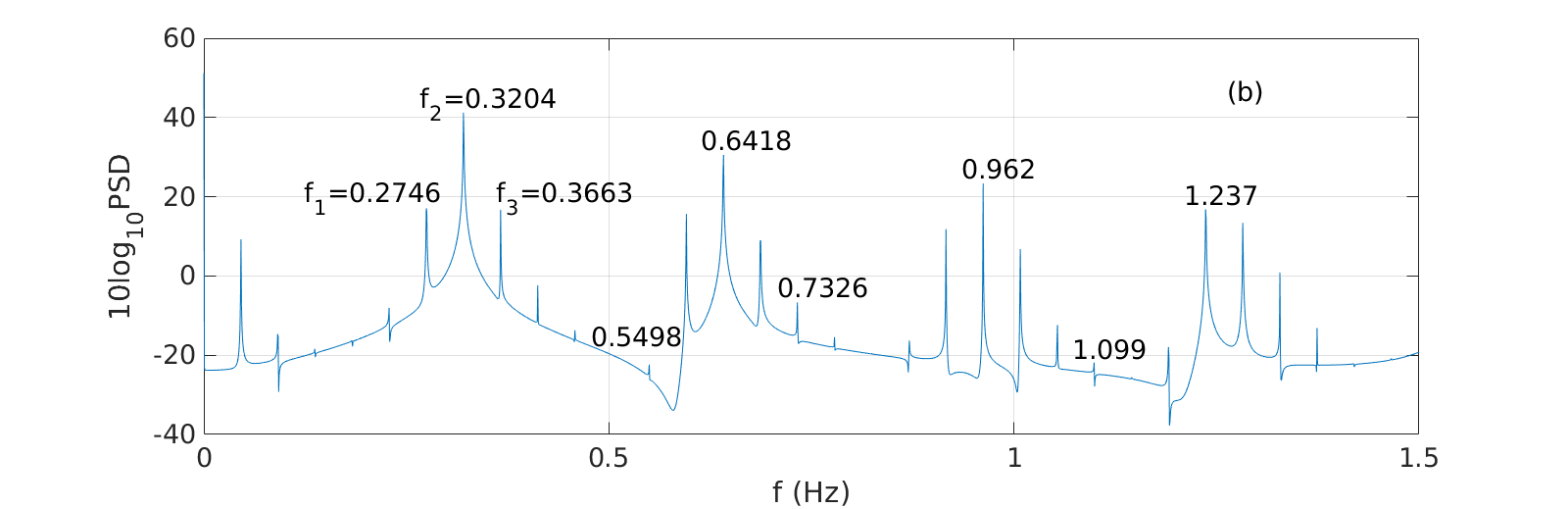}
	\includegraphics[scale=0.22, angle=0]{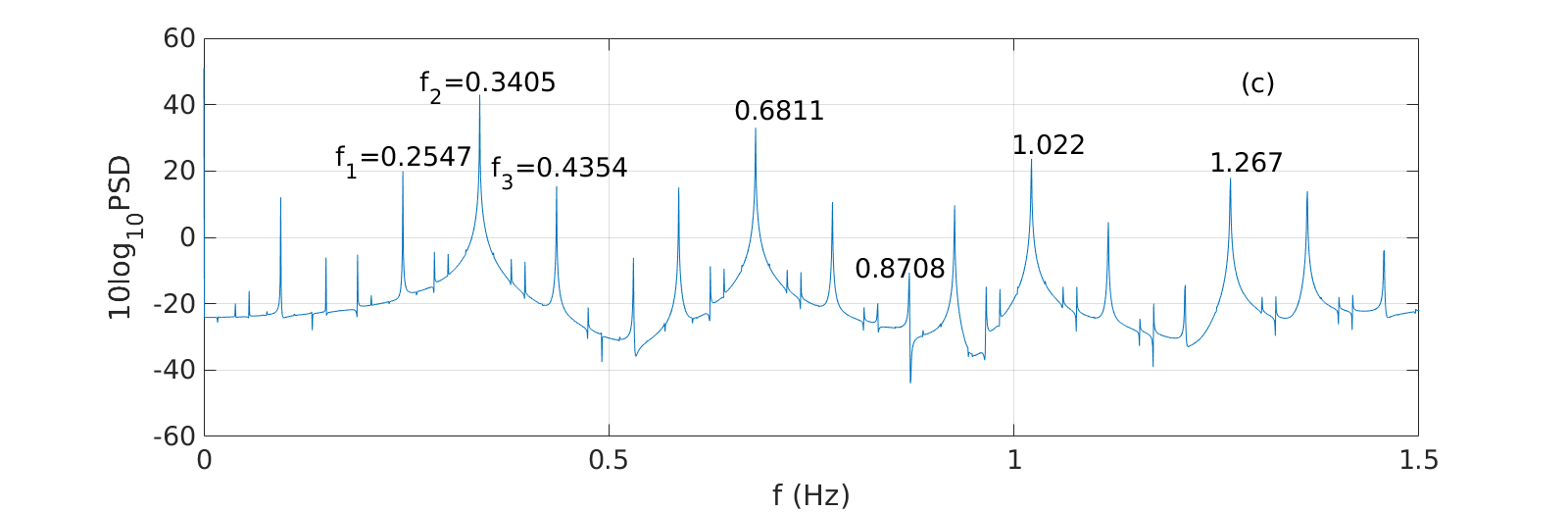}
	\includegraphics[scale=0.22, angle=0]{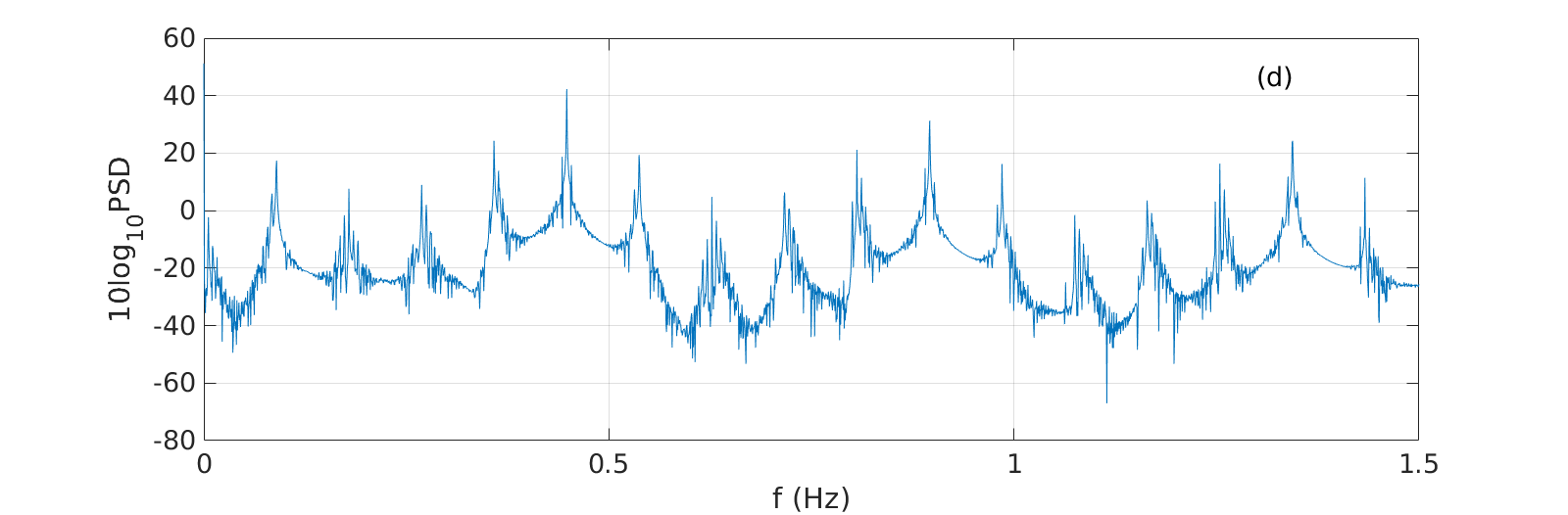}
	\includegraphics[scale=0.22, angle=0]{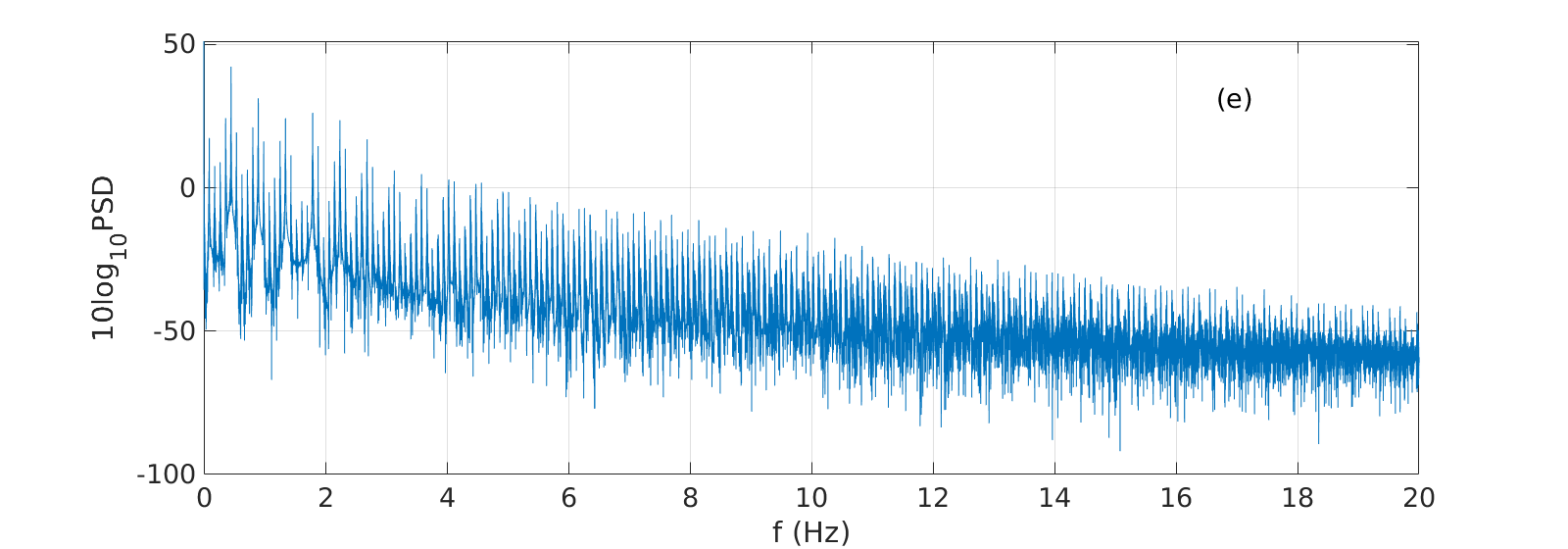}
	\caption{(Color online)  PSD for $E=24,\,y=1.0,\, p_{y}>0,\,K_{x}=26.75,\, K_{y}=26.75,\,x_{c}=1.1,\,y_{c}=1.0$ and for different values of accelerations $a$, namely $(a)\,a=0.20,\,(b)\,a=0.27,\,(c)\,a=0.295,\,(d)\,a=0.35$. Upto $a=0.27$, only $f_{1},f_{2}$ and $f_{3}$ and their harmonics are present. From $a=0.295$ onwards more frequencies start populating the spectrum. At $a=0.362$, the frequencies are highly populated (see Fig.$\ref{fig:PSD_E24}(e)$ which indicates the onset of chaos. }
	\label{fig:PSD_E24}
\end{figure}

Next we plot the PSD diagrams for different values of the acceleration $(a=0.20,0.27,0.295\,\text{and}\,0.362)$ of the system for a fixed value of energy $E=24.0$ (see Fig. \ref{fig:PSD_E24}). For lower value of the particle acceleration i.e at $a=0.20$ the PSD (Fig. \ref{fig:PSD_E24}($a$)) shows the presence of only three frequencies, which are $f_{1}=0.1979, f_{2}=0.2256,\,f_{3}=0.2538$ and their harmonics. While for larger values of $a$ that is at $a=0.27$ (Fig. \ref{fig:PSD_E24}($b$)) and at $a=0.295$ (Fig. \ref{fig:PSD_E24}($c$)) the frequencies $f_{1},f_{2}\,\text{and}f_{3}$ start getting shifted and other frequencies near $f_{1},f_{2}\,\text{and}f_{3}$ start getting populated like the previous case. Finally, for very large value of $a=0.362$ (Fig. \ref{fig:PSD_E24}($d$)) where the interaction between the harmonic potential and the horizon is very large, we found that the PSD diagrams are fully populated with again an exponential decay over the entire frequency region which basically shows the distinguishing feature of the onset of chaos. Hence we obtain a series of transitions in the dynamics of our composite system, namely, from periodic to quasi-periodic and finally to chaotic dynamics as the total energy of the system $(E)$ or the acceleration of the particle $(a)$ is increased. Thus both the cases result in \textit{ quasi-periodic route to chaos}.

\subsection{\label{LE}Lyapunov exponent}
In literature the Lyapunov exponent of a dynamical system is defined as the quantity that characterizes the rate of separation of infinitesimally close trajectories \cite{sandri:96}. In phase space if we consider two trajectories with initial separation 
$\delta x_{0}$, the rate of divergence within the linearized approximation is given by 
\begin{eqnarray}
|\delta x(\lambda)|\approx e^{\lambda_{L}\lambda}|\delta x_{0}|
\end{eqnarray}
where $\lambda_{L}$ is the Lyapunov exponent. In this situation, it may be worth to point out that the value of the Lyapunov exponent has an upper bound which was first mentioned in \cite{Maldacena:2015waa} for the SYK model. For our system the Lyapunov exponent $(\lambda_{L})$ is bounded by
\begin{eqnarray}
\lambda_{L}\leq a~,
\label{R1}
\end{eqnarray} 
where $a$ is the acceleration of the Rindler frame. This is apparent from the near horizon solution (\ref{1.08}) for $x$. Now at the quantum level, the accelerated frame will feel the Unruh temperature $T=\hbar a/2\pi$ (see \cite{Unruh:1976db}) and hence the above inequality can be expressed as $\lambda_L\leq(2\pi T)/\hbar$, which is exactly identical to what was obtained for SYK model \cite{Maldacena:2015waa}. In this regard we here mention some recent studies on the bound on Lyapunov exponents in the context of black holes. In \cite{Lu:2018mpr} authors have studied the minimal length effects on the chaotic motion of a massive particle near the black hole horizon where they have shown that the corresponding Lyapunov exponent is greater than that in the usual case with the absence of the minimal length using the Hamilton-Jacobi method. People have also shown that the bound on the Lyapunov exponent can be violated by a large number of black holes including the RN-dS black holes or black holes in Einstein-Maxwell-Dilaton, Einstein-Born-Infeld and Einstein-Gauss-Bonnet-Maxwell gravities \cite{Zhao:2018wkl}. Another interesting work \cite{Cubrovic:2019qee} where the author performed a systematic study of the maximum Lyapunov exponent values for the motion of classical closed strings in Anti-de Sitter black hole geometries with spherical, planar and hyperbolic horizons. The analytical estimation from the linearized variational equations predict the Lyapunov exponent value modified as $\lambda_L \approx 2\pi T n$, where $n$ is the winding number of the string.

\begin{figure}[!ht]
	\centering
	\includegraphics[scale=0.35, angle=0]{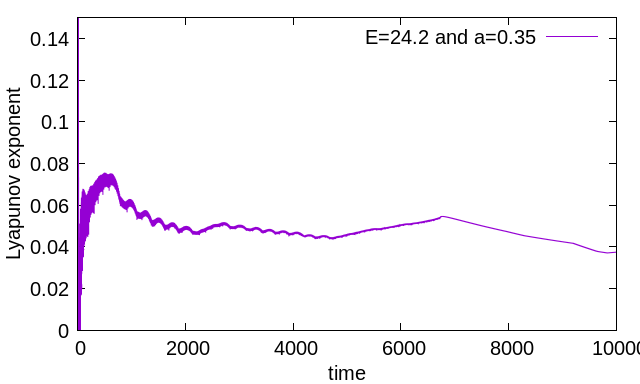}
	\caption{(Color online) Largest Lyapunov exponent for the particle in accelerated frame at the energy value $E=24.2$ and the acceleration $a=0.35$. The exponent settles at positive value $\sim 0.06$ which is lower than the upper bound (0.35).  }
	\label{fig:lyp1}
\end{figure}

\begin{figure}[!ht]
	\centering
	\includegraphics[scale=0.35, angle=0]{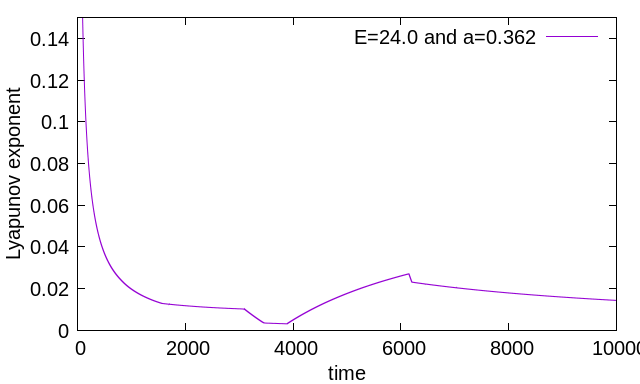}
	\caption{(Color online) Largest Lyapunov exponent for the particle in accelerated frame at the energy value $E=24$ and the acceleration $a=0.362$. The exponent settles at positive value $\sim 0.02$ which is lower than the upper bound (0.362).  }
	\label{fig:lyp2}
\end{figure}

Now below we present our numerical analysis on Lyapunov exponents for different cases in order to show our proposition is reasonable. We have plotted the largest Lyapunov exponent for two cases. First, we consider the energy $E=24.2$ and $a=0.35$ (see Fig. \ref{fig:lyp1}) and secondly, we have plotted for $E=24.0$ and $a=0.362$ (see Fig. \ref{fig:lyp2}). For both the cases we obtained the chaotic behaviour in the particle dynamics. In both the figures (Fig. \ref{fig:lyp1} and Fig. \ref{fig:lyp2}) it is observed that the Lyapunov exponent settle to positive values ($\sim 0.04$ and $\sim 0.02$ respectively) which suggests the chaotic motion of the particle. Since we observed that the obtained values of the Lyapunov exponents for both the cases are lower than the bounds (0.35 and 0.362 respectively), it is consistent with our claim. 

\noindent

\section{\label{DC}Discussions and conclusions}
In this paper we have studied the motion of a massless and chargeless particle in an accelerated frame in the flat spacetime background. First we derived the equations of motion of the outgoing particle trajectories and found that the position coordinate along a particular direction increases exponentially with the time parameter which gives us the indication that may be the horizon is influencing the chaotic fluctuation in the particle's motion. Next our aim was to trap the particle near the horizon and study its dynamical behaviour. For this purpose we introduced an external harmonic potential in the accelerated frame and derived the equations of motion from the Hamiltonian of the whole composite system. The whole phenomena are seen by a comoving observer. After solving these equations numerically and analyzing the Poincar$\acute{e}$ sections and the PSD diagrams we ascertain our claim that chaos can be indeed induced in the particle dynamics by just influence of a horizon.

So it appears that  a harmonic oscillator in the Rindler frame lost its  periodicity with respect to the comoving observer as it  approaches nearer to the horizon. This is consistent with the well known KAM theory. According of which, if an integrable system having the number of degrees of freedom (say $n$) greater than three, it loses its stability with the application of a small perturbation into the system \cite{Kaloshin}. Now our system has $n=4$ degrees of freedom (namely, $x,p_x,y,p_y$) and that means a general small perturbation can lead our system to instability which is happening for our case. Here if we think the interaction with the horizon as a perturbation then as the perturbation gets increased (due to increase in $E$ and $a$) the periodic motion of the harmonic oscillator initially decomposes into a number of tori. Further increase in the parameters $(E$ and $a)$ leads to the instability in some of the regular tori that results them breaking into several small tori. It signifies that our harmonic oscillator initially started with a fundamental frequency but later due to the increment in the perturbation the fundamental frequency decomposes and a number of other frequencies show up into the system which is the characteristics of the onset of chaos. Now from the expression of the Hamiltonian of the system we can say that the first term of Eq. (\ref{1.09}) will act as the perturbation in the system and  the other two  terms in Eq. (\ref{1.09}) will form the integrable system of harmonic oscillator in two dimensions. The only source of this perturbation term is the acceleration of the system itself. That means if an harmonic oscillator is moving in two dimensions in a constant accelerated frame, for a comoving observer it would appear that the acceleration itself perturbed the  system rigorously and as a result the harmonic oscillator loses its fundamental frequency and for large perturbation it shows chaotic behavior.  

Chaos in the coupled harmonic oscillators (as for an example, double pendulum) system in the flat space time for inertial observer is generally observed due to the presence of the nonlinear coupling \cite{Nbook}. The coupling is independent of the state of the inertial observer. So chaos appears in the system even the velocity of the inertial observer is set to zero. Here, as a precautionary note for the readers, we would like to mention that in our system the coupling between the oscillators appears owing to the presence of the accelerated observer only which in turn provides a Killing horizon. The coupling among the oscillators are happening through their interaction with this horizon. In absence of the acceleration, the coupling between the oscillator vanishes and that will lead to the disappearance of the chaos from the system. So one needs to be careful while comparing the chaos that appears in our model to those appear in the system of coupled harmonic oscillators for inertial observer.

We also found that the chaos of the particle probing the Rindler horizon has a universal upper bound on the Lyapunov exponent.  In this regard, it may be mentioned that the upper bound is given by the acceleration of the particle and it is independent of the external potentials which prevents the particles falling into the horizon (for details on the potential independent nature, see \cite{Hashimoto:2016dfz}).  This upper bound conjecture of the Lyapunov exponent coincides with the prediction of the SYK model, argumented by Maldacena, Shenkar and Stanford \cite{Maldacena:2015waa}.

From this analysis it is apparent that an integrable system when comes under the influence of horizon, it's dynamics can be chaotic. Since in the present case, the horizon is part of the flat background (vanishing Riemann curvature), the conjecture -- horizon makes system chaotic -- is further balustrade as the curvature does not play much role. This has been further elaborated by studying the route to chaos for our present model and the model with harmonic oscillator in static spherically symmetric (SSS) black hole. We observed that they are similar in nature (see Appendix \ref{Appendix2} for the PSD of SSS case). 

The present analysis is well consistent with the equivalence principle as only horizon is responsible for similar chaotic dynamics in both the setups: accelerated frame and black hole spacetime. In this regard we further want to point out a possible parallel description between the  thermality in an accelerated frame (known as Unruh effect \cite{Unruh:1976db}) and that for a static observer in black hole  spacetime (familiar as Hawking effect \cite{Hawking:1974rv}). It is well known that both the temperature expressions are identical with the identification of the acceleration parameter as the surface gravity for black hole. Similar situation again arises here also. It is noticeable that the maximum value of Lyapunov exponent is given by proper acceleration of the frame (see Eq. (\ref{R1})). Whereas, it was already mentioned in \cite{Dalui:2018qqv} that the same for black hole case is given by the surface gravity. So the existing map between acceleration and surface gravity at the quantum level also appears here, but at the classical level.

Furthermore, we hope that as our model composed of simple accelerated frame and harmonic oscillator, an experimental setup may be designed to test this possibility. In addition we point out that in the present context the Rindler spacetime was investigated in \cite{Hashimoto:2016dfz} for a massive particle in original Rindler coordinates. The bound was obtained same as ours (\ref{R1}). Here we worked out for a chargeless and massless particle and the Rindler background has been taken in Painleave coordinates. Interestingly, in both analysis the bound on $\lambda_L$ is identical and independent of the mass of the particle. This certainly provides an universal feature of horizon as mass of the particle does not play any role on this bound. In this sense, the current analysis is different from the earlier discussions \cite{Bombelli:1991eg,Sota:1995ms,Vieira:1996zf,  Suzuki:1996gm,Cornish:1996ri,deMoura:1999wf,Hartl:2002ig,Han:2008zzf,Takahashi:2008zh,Hashimoto:2016dfz,Li:2018wtz} (as they either used curved background or massive particle or both simultaneously in their analysis). In this paper we used massless particle and also investigated the nature of power spectrum density which gives the information on the nature of {\it route to chaos} when one changes the system parameters. This has not been attempted in earlier analysis. In appendix \ref{Appendix1} we also give the PSD for the massive particle in Rindler frame. We found that route to chaos for both massless and massive is similar in nature -- frequency spectrum gets populated as one increases energy or acceleration. This again shows that the horizon actually can make system chaotic while the mass of the particle is not so important.

Let us now summarize what are the information we can have from the study of PSD of our system.
\begin{itemize}
\item Study of PSD helps to understand the actual appearance of chaos in a system by analyzing the population of the spectrum with the change of macroscopic parameters of the system. 
\item Here we studied the motion of massless as well as massive particle in a Rindler frame. In both cases, after analyzing PSDs, we observed that the nature of onset of chaos into the system is similar. Interestingly, the system with massive and massless particle exhibit same route to chaos which in the present case we identified as quasi-periodic route to chaos. Increase of same parameters (e.g. energy and acceleration) for both the situations, frequency spectra get populated. So we infer that the horizon actually induces chaos to the system, while mass of the particle is not so important.
\item At the stage of maximum chaos, the PSDs show an exponential decay in all cases. Although as of now this is completely a phenomenological feature, but it may be possible that such nature is a very general one for any temporal chaos in a system. This statement is just a prediction rather than a concrete one.
\end{itemize}

Finally, in this scenario where researchers are trying to prove the equivalence principle experimentally, we hope our work can shed some light in that direction. A significant amount of efforts have been put to uncover the universal behaviour of the Lyapunov exponent and relate it with the temperature of the chaotic system. 
With the progress of these present investigations we think that the mystery of horizon may be revealed and further exploration on the behaviour of the universal chaos will benefit the physics of black holes and quantum information.    

\appendix
\begin{widetext}
\section{\label{Appendix1}Presence of chaos in an accelerated frame for a massive particle}
The expression of the total energy of the composite system for a massive test particle moving in the background (\ref{1.03}) under the influence of the two harmonic oscillators is 
\begin{eqnarray}
E=&&-\sqrt{1-2ax}~p_x +\sqrt{p_x^2+p_y^2+m^2}+\f{1}{2}K_{x}(x-x_{c})^{2}+\f{1}{2}K_{y}(y-y_{c})^{2}~,\label{App1}
\end{eqnarray}
and the corresponding equations of motion are
\begin{eqnarray}
&&\dot{x}=\f{\p E}{\p p_{x}}=-\sqrt{1-2ax}+\f{p_{x}}{\sqrt{p_{x}^{2}+p_{y}^{2}+m^2}}~;\label{App2}\\
&&\dot{p_{x}}=-\f{\p E}{\p x}=-\f{a}{\sqrt{1-2ax}}p_{x}-K_{x}(x-x_{c})~;\label{App3}\\
&&\dot{y}=\f{\p E}{\p p_{y}}=\f{p_{y}}{\sqrt{p_{x}^{2}+p_{y}^{2}+m^2}}~;\label{App4}\\
&&\dot{p_{y}}=-\f{\p E}{\p y}=-K_{y}(y-y_{c})~.\label{App5}
\end{eqnarray}    
\subsection{Poincar$\acute{e}$ sections for constant total energy of the system}
Here we present the Poincar$\acute{e}$ sections for a massive particle moving with an acceleration in flat spcaetime for a fixed energy value of the system $(E=24.0)$. The dynamical equations (Eqs. (\ref{App2})-(\ref{App5})) are solved  numerically using the fourth order Runge-Kutta method with fixed $dt=5\times10^{-3}$. The mass of the particle is chosen $m=1$. The other parameters are kept same as mentioned in the section~\ref{NA}. The analysis confirms that in presence of horizon with the increase in the particle acceleration the motion of the massive particle becomes chaotic (see Fig. \ref{fig:Poinc_E24_massive}) just like the earlier case i.e for the massless particle.  
\begin{figure}[!ht]
	\centering
	\includegraphics[scale=0.30, angle=0]{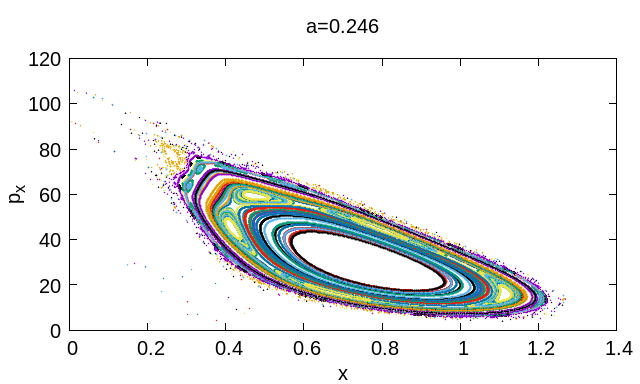}
	\includegraphics[scale=0.30, angle=0]{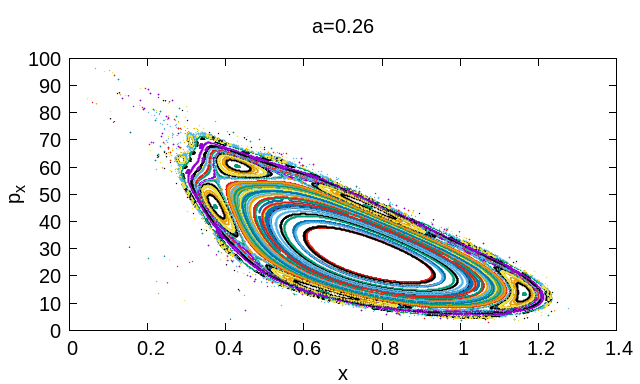}
	\includegraphics[scale=0.30, angle=0]{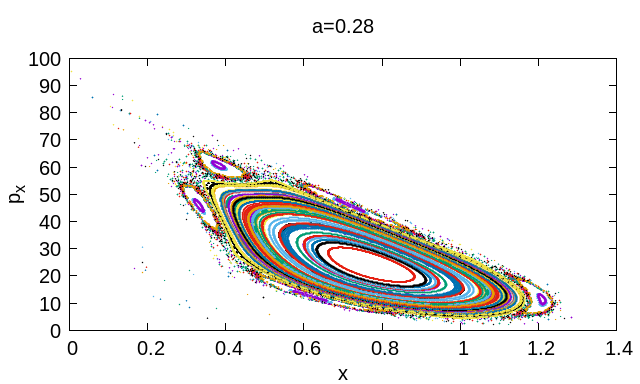}
	\includegraphics[scale=0.30, angle=0]{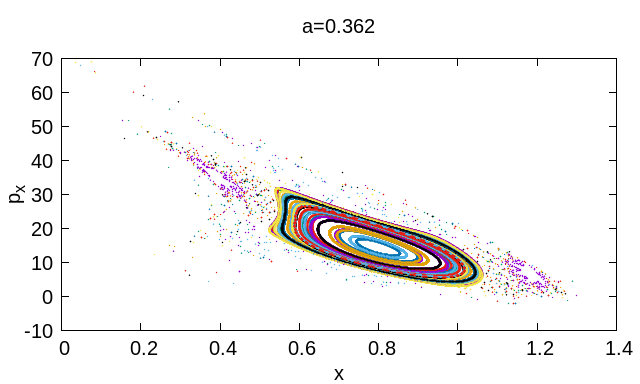}
	\caption{(Color online) The Poincar$\acute{e}$ sections in the $(x,p_{x})$ plane with $y=1.0$ and $p_{y}>0$ at different values of acceleration of the system with mass $m=1$ for fixed energy ($E=24.0$). The values of accelerations are $a=0.246, 0.26, 0.28$ and $0.362$. The other parameters are $K_{x}=26.75,\, K_{y}=26.75,\,x_{c}=1.1$ and $y_{c}=1.0$. For large value of acceleration the KAM Tori break and the scattered points emerge which indicates the onset of chaotic dynamics.   }
	\label{fig:Poinc_E24_massive}
\end{figure}

\subsection{Poincar$\acute{e}$ sections for constant acceleration of the system}
In this part we present the Poincar$\acute{e}$ sections for the massive particle confined in a two dimensional harmonic potential moving with a fixed acceleration $(a=0.35)$ in flat spcaetime but for different values of energy of the system. The numerical simulations are done in the similar process as earlier keeping all the values of the parameters same. The analysis shows us that in presence of horizon with the increase in the total energy of the composite system the motion of the massive particle becomes chaotic which is quite evident from the following figures (see Fig. \ref{fig:Poinc_a0p35_massive}).  
\begin{figure}[!ht]
	\centering
	\includegraphics[scale=0.30, angle=0]{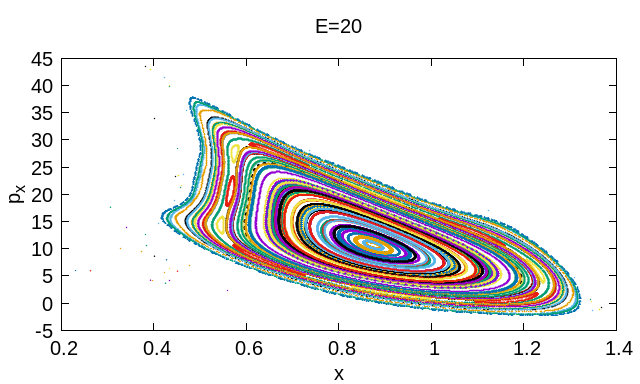}
	\includegraphics[scale=0.30, angle=0]{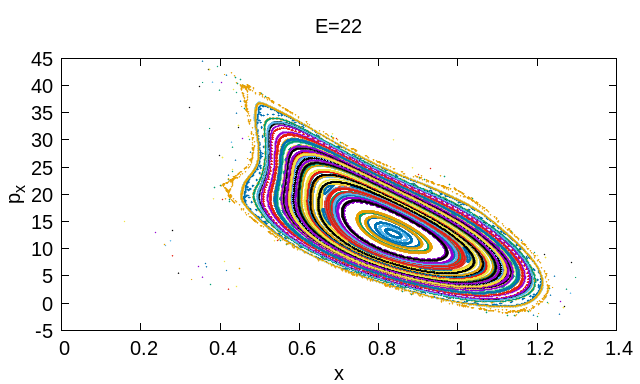}
	\includegraphics[scale=0.30, angle=0]{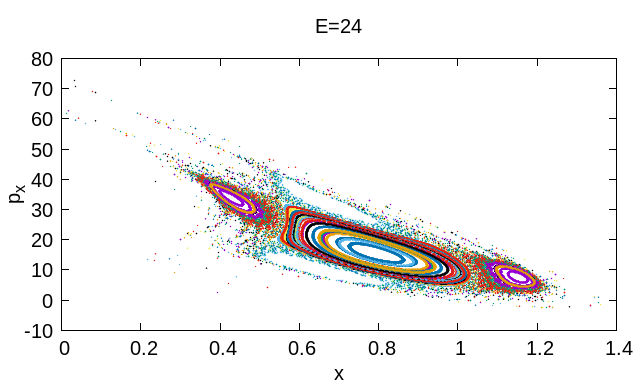}
	\includegraphics[scale=0.30, angle=0]{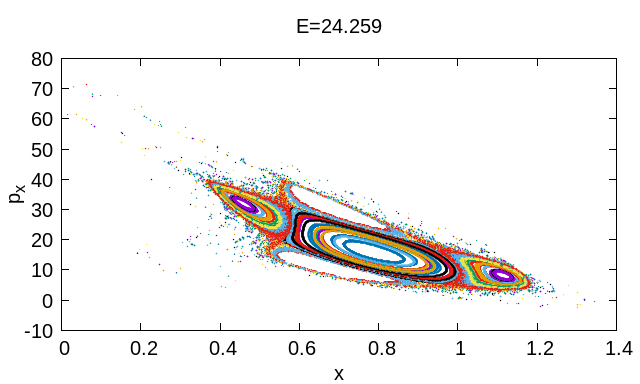}
	\caption{(Color online) The Poincar$\acute{e}$ sections in the $(x,p_{x})$ plane with $y=1.0$ and $p_{y}>0$ at different values of energy of the system with mass $m=1$ but for fixed acceleration ($a=0.35$). The energies are $E=20,22,24$ and $24.259$. The other parameters are $K_{x}=26.75,\, K_{y}=26.75,\,x_{c}=1.1$ and $y_{c}=1.0$. For large value of energy the KAM Tori starts breaking and the scattered points begin to appear which indicates the presence of chaotic motion in the particle dynamics.}
	\label{fig:Poinc_a0p35_massive}
\end{figure}

\subsection{PSD for a massive particle for constant total energy of the system}
In this part we present the PSD diagrams for the massive particle confined in a two dimensional harmonic potential moving with a constant total energy of the system $(E=24)$ in flat spcaetime but for different values of accelerations. The numerical simulations are done in the similar process as described earlier (see \ref{PSD}) keeping all the values of the parameters same. The analysis shows us that in presence of horizon with the increase in the acceleration of the system, the motion of the massive particle becomes chaotic which is clearly observed as the frequencies start populating the spectrum. From the following figures (see Fig. \ref{fig:PSD_E24_m1}) it can be seen for high values of accelerations the frequencies are highly populated which indicates the beginning of the chaotic fluctuations into the system. 
\begin{figure}[!ht]
	\centering
	\includegraphics[scale=0.176, angle=0]{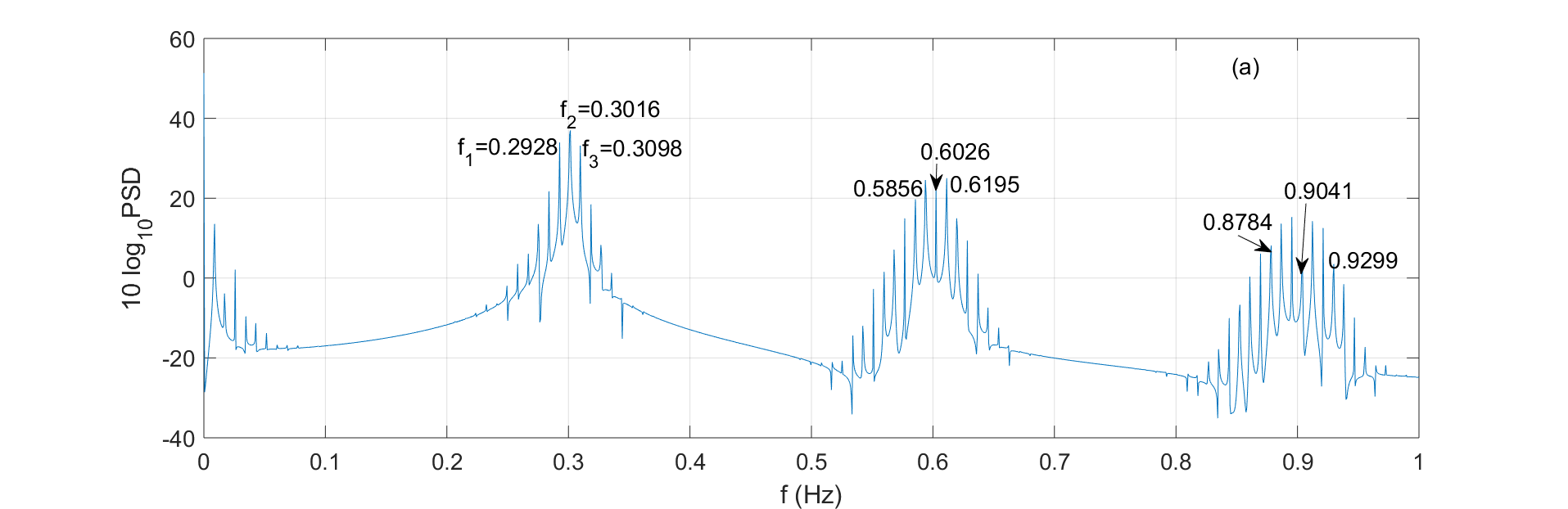}
	\includegraphics[scale=0.1755, angle=0]{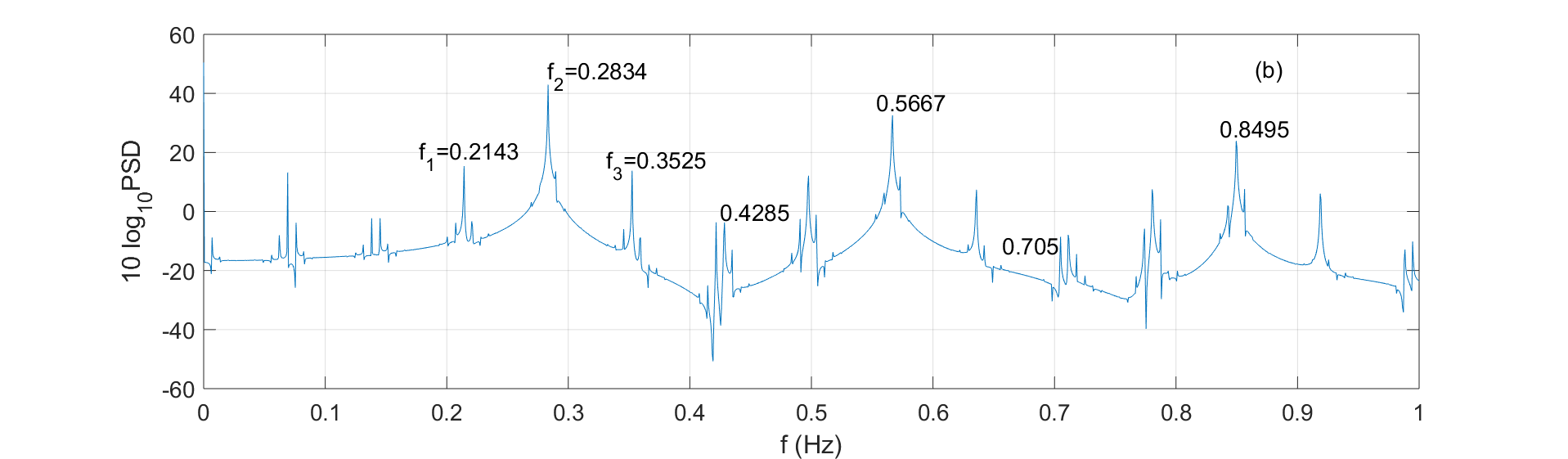}
	\includegraphics[scale=0.176, angle=0]{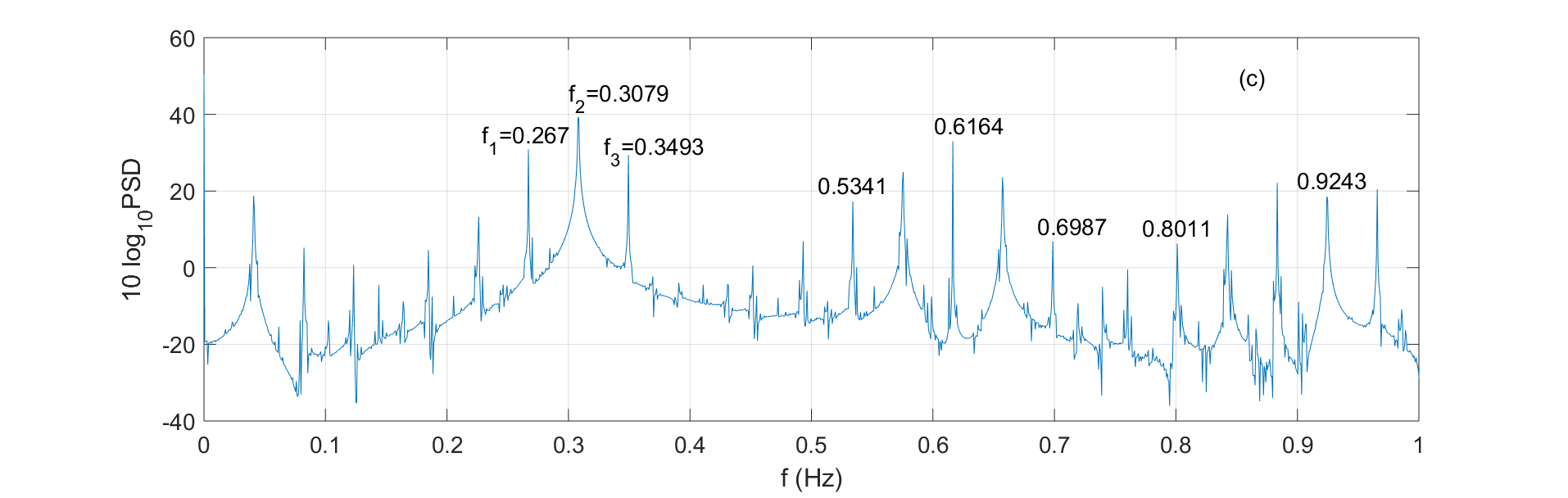}
	\includegraphics[scale=0.1755, angle=0]{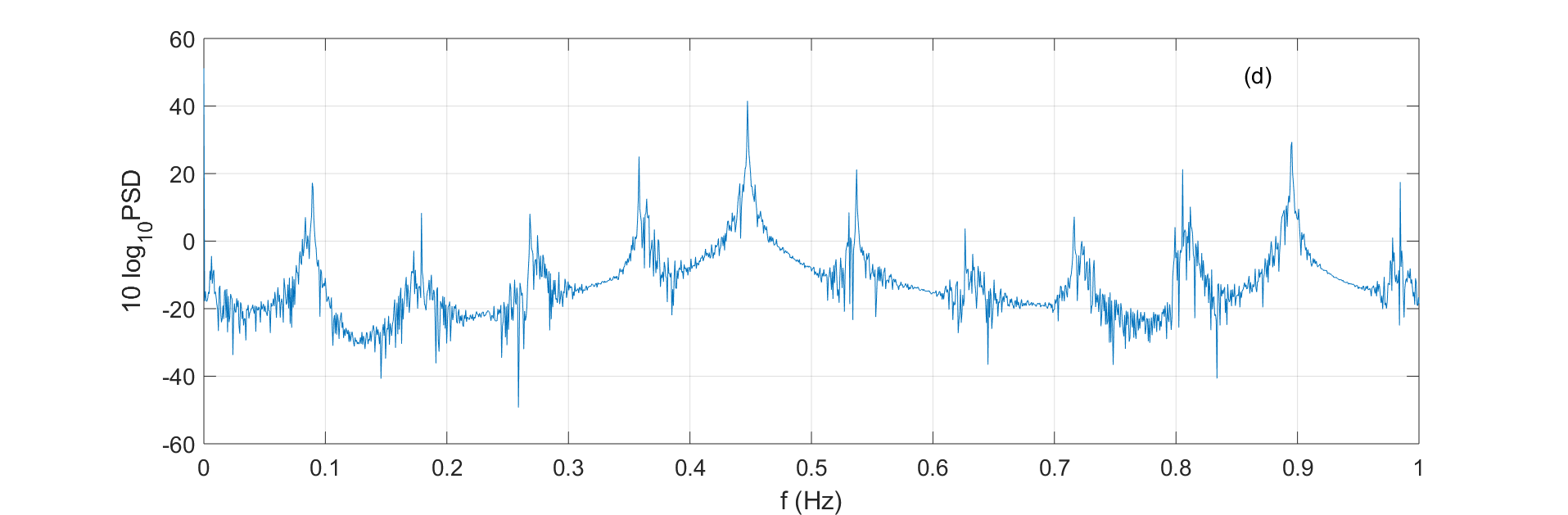}
	\includegraphics[scale=0.18, angle=0]{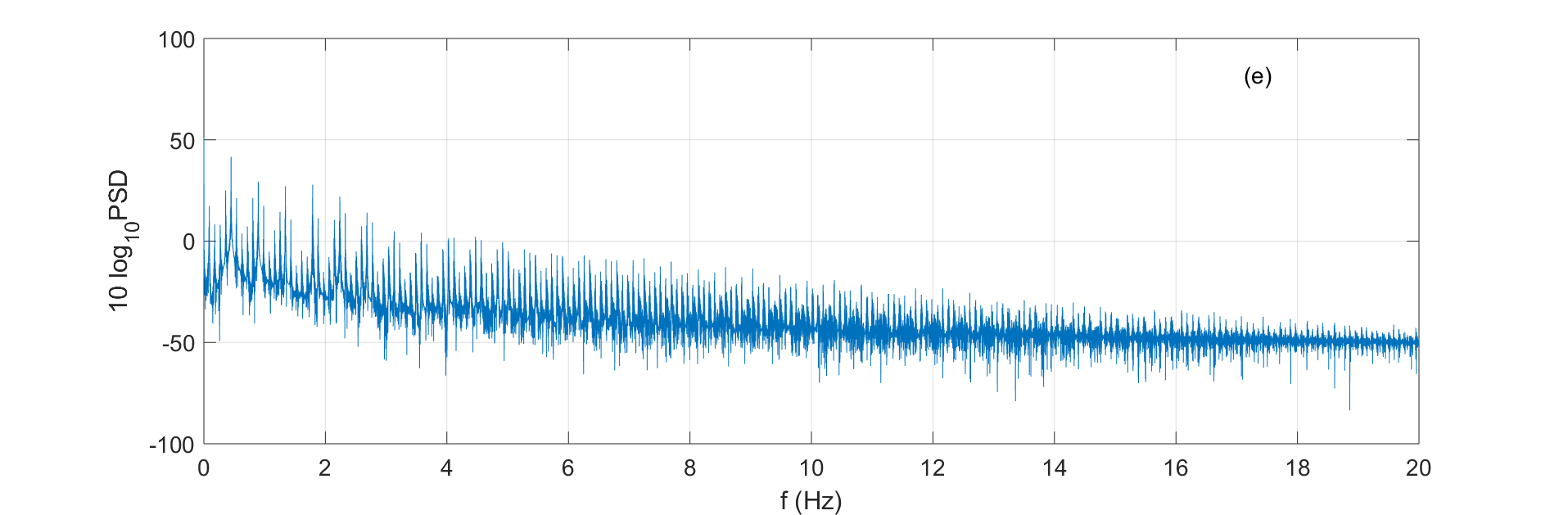}
	\caption{(Color online) PSD for a massive particle with mass $m=1,\,E=24,\,y=1.0,\, p_{y}>0,\,K_{x}=26.75,\, K_{y}=26.75,\,x_{c}=1.1,\,y_{c}=1.0$ and for different values of accelerations $a$, namely $(a)\,a=0.246,\,(b)\,a=0.26,\,(c)\,a=0.28,\,(d)\,a=0.362$. From the figure it can be seen that from $a=0.28$ onwards more frequencies start populating the spectrum. At $a=0.362$, the frequencies are highly populated (see Fig.$\ref{fig:PSD_E24_m1}(e)$ which indicates the onset of chaos.}\label{fig:PSD_E24_m1}
\end{figure}

\subsection{PSD for a massive particle for constant acceleration of the system}
\begin{figure}[!ht]
	\centering
	\includegraphics[scale=0.176, angle=0]{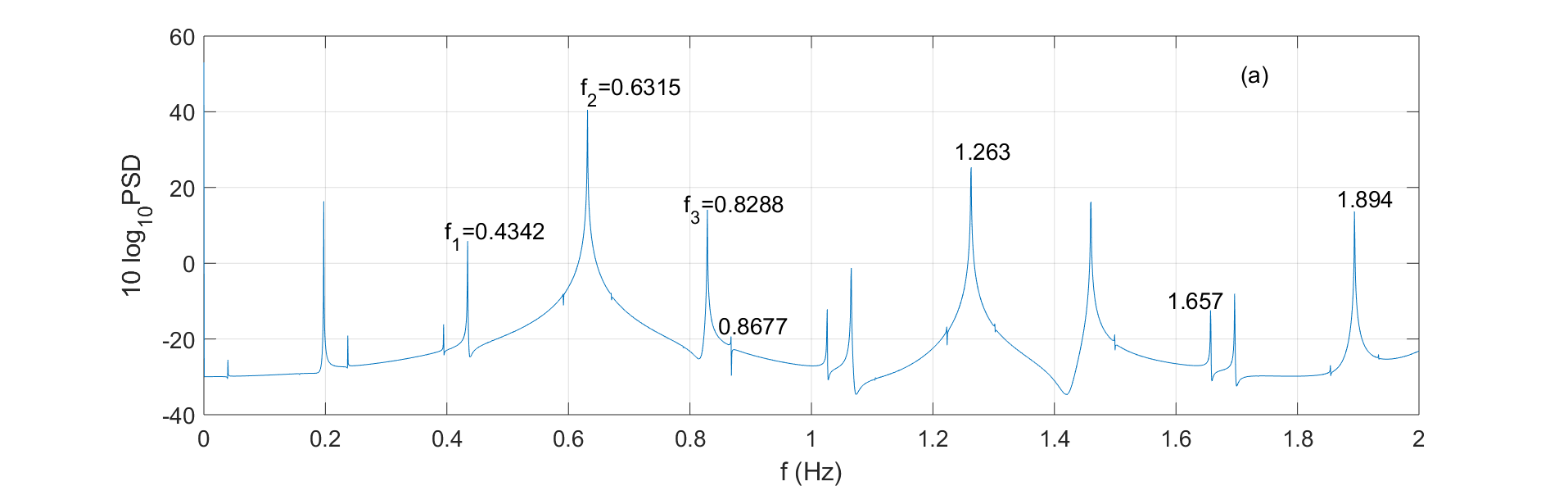}
	\includegraphics[scale=0.1755, angle=0]{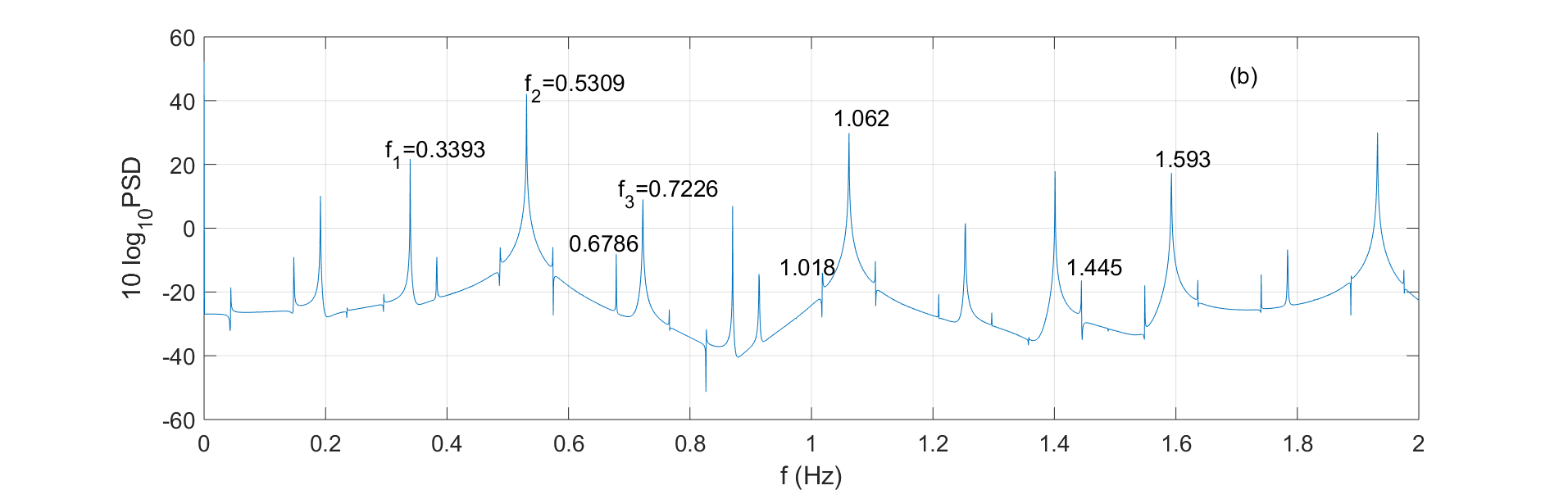}
	\includegraphics[scale=0.176, angle=0]{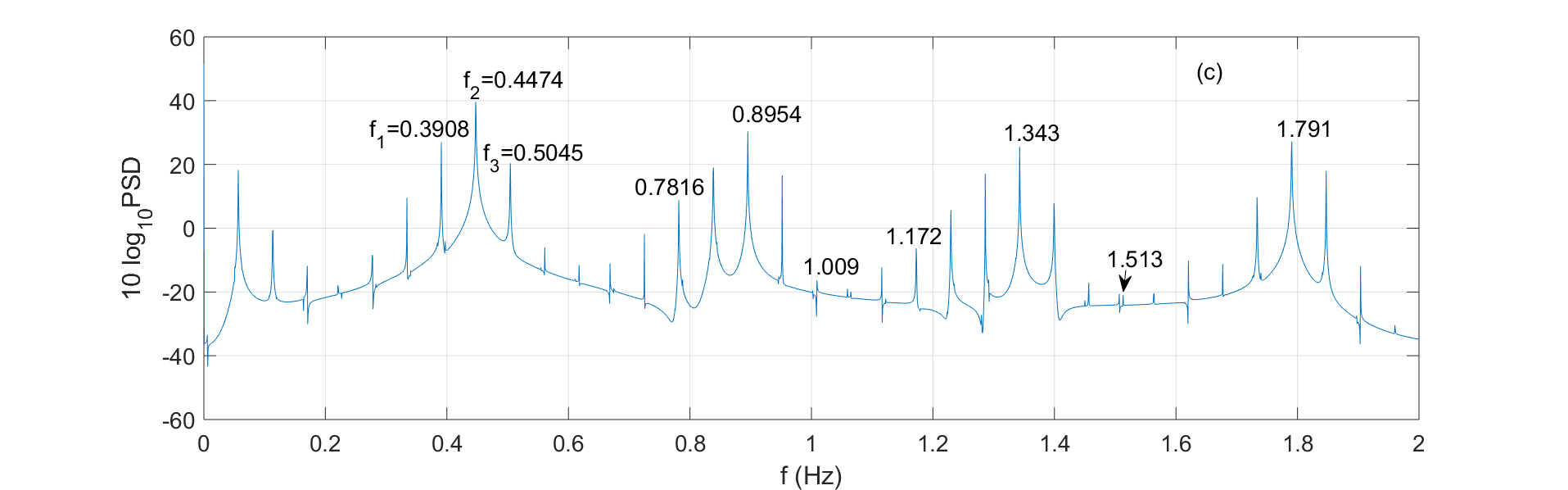}
	\includegraphics[scale=0.1755, angle=0]{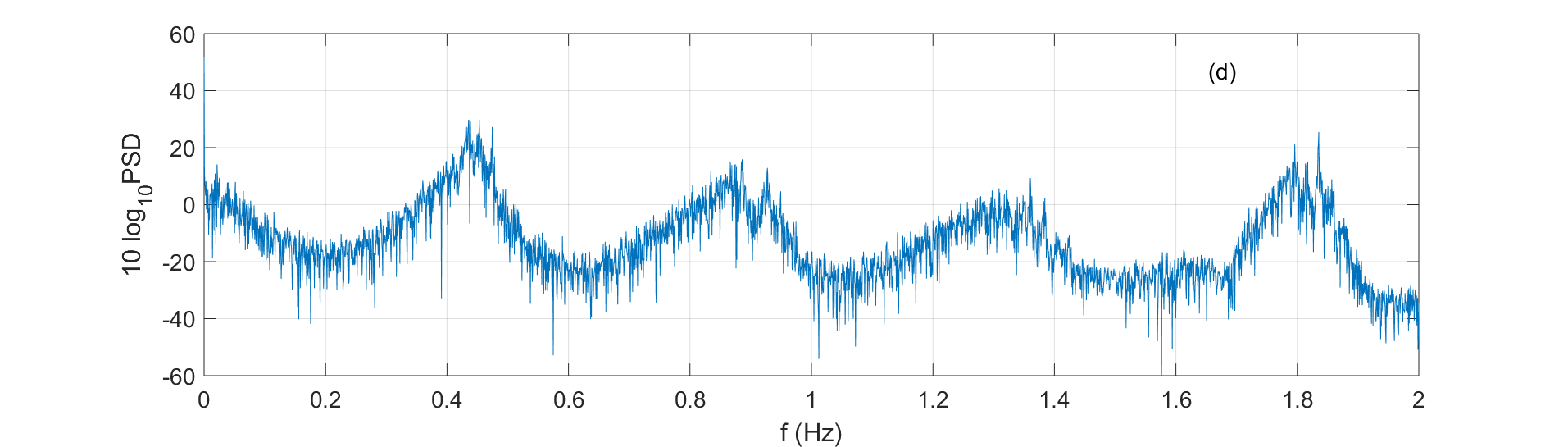}
	\includegraphics[scale=0.18, angle=0]{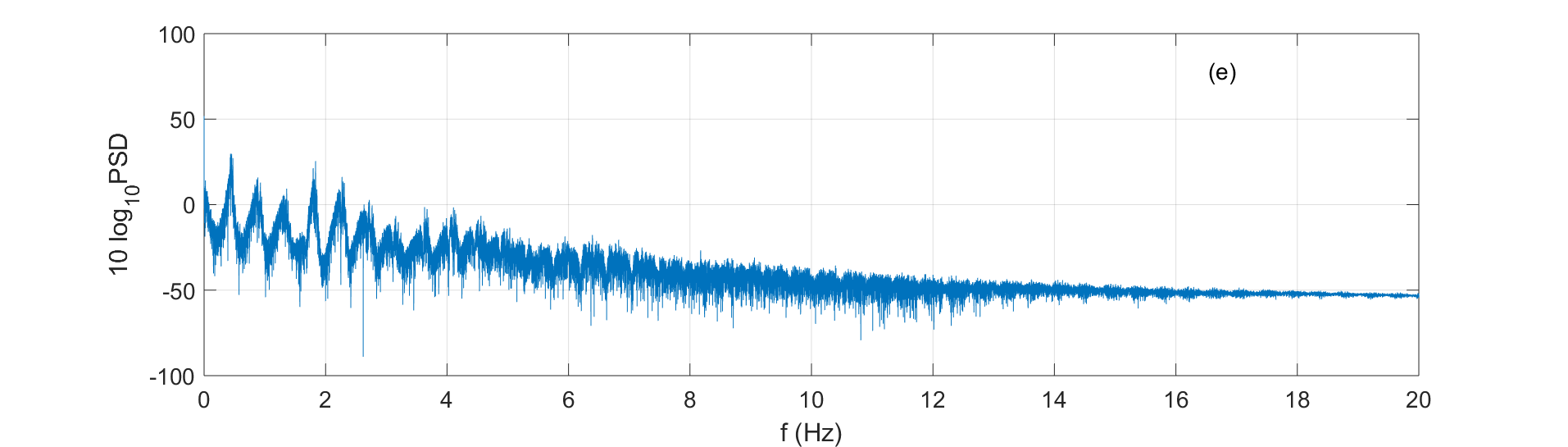}
	\caption{(Color online) PSD for a massive particle with mass $m=1,\,a=0.35,\,y=1.0,\, p_{y}>0,\,K_{x}=26.75,\, K_{y}=26.75,\,x_{c}=1.1,\,y_{c}=1.0$ and for different values of energies $E$, namely $(a)\,E=20,\,(b)\,E=22,\,(c)\,E=24,\,(d)\,E=24.259$. From the figure it can be seen that from $E=24$ onwards more frequencies start populating the spectrum. At $E=24.259$, the frequencies are highly populated (see Fig.$\ref{fig:PSD_a0p35_m1}(e)$ which indicates the onset of chaos.}\label{fig:PSD_a0p35_m1}
\end{figure}	
Here we present the PSD diagrams for the massive particle confined in a two dimensional harmonic potential which is moving with a constant acceleration $(a=0.35)$ in flat spacetime but for different values of the total energy of the system. The numerical simulations are done in the similar process as earlier keeping all the values of the other parameters same. This analysis shows us that in presence of horizon with the increase in the total energy of the system, the motion of the massive particle becomes chaotic and it is observed as the frequencies start populating the spectrum as the value of the total energy goes higher. From the following figures (see Fig. \ref{fig:PSD_a0p35_m1}) it can be seen for high value of the energy (Fig. \ref{fig:PSD_a0p35_m1}$(e)$) the frequencies are highly populated which indicates the beginning of the chaotic fluctuations into the system.

\subsection{Lyapunov exponent for massive particle}
In this section we present our numerical analysis on Lyapunov exponents just like the earlier one (see \ref{LE}) but for the massive particle with mass $m=1$. We have plotted the largest Lyapunov exponent for two cases. First, we consider the energy $E=24.259$ and $a=0.35$ (see Fig. \ref{fig:lyp1m1}) and secondly, we have plotted for $E=24.0$ and $a=0.362$ (see Fig. \ref{fig:lyp2m1}). For both the cases we obtained the chaotic behaviour in the particle dynamics. In both the figures (Fig. \ref{fig:lyp1m1} and Fig. \ref{fig:lyp2m1}) it is observed that the Lyapunov exponent settle to positive values ($\sim 0.007$ and $\sim 0.033$ respectively) which suggests the chaotic motion of the particle. Since we observed that the obtained values of the Lyapunov exponents for both the cases are lower than the upper bounds (0.35 and 0.362 respectively), it is consistent with our claim for the massive case also. 
\begin{figure}[!ht]
	\centering
	\includegraphics[scale=0.40, angle=0]{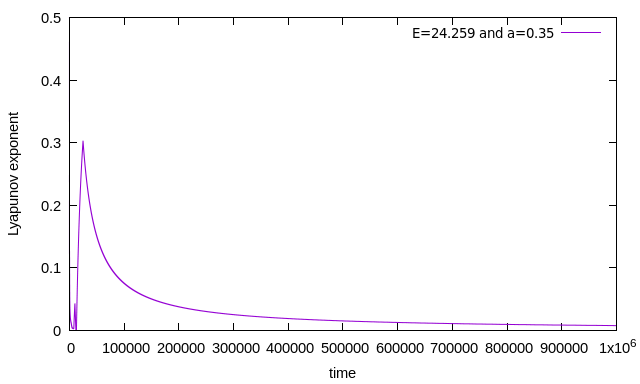}
	\caption{(Color online) Largest Lyapunov exponent for the particle in accelerated frame at the energy value $E=24.259$ and the acceleration $a=0.35$. The exponent settles at positive value $\sim 0.007$ which is lower than the upper bound (0.35).  }
	\label{fig:lyp1m1}
\end{figure}
\begin{figure}[!ht]
	\centering
	\includegraphics[scale=0.40, angle=0]{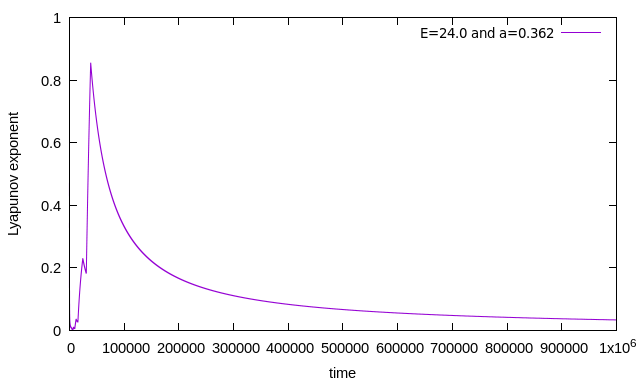}
	\caption{(Color online) Largest Lyapunov exponent for the particle in accelerated frame at the energy value $E=24$ and the acceleration $a=0.362$. The exponent settles at positive value $\sim 0.033$ which is lower than the upper bound (0.362).  }
	\label{fig:lyp2m1}
\end{figure}
\noindent

\section{\label{Appendix2}Power Spectral Density analysis for a system of simple harmonic motions of a particle in a static spherically symmetric (SSS) spacetime}
In our previous analysis \cite{Dalui:2018qqv} we studied a system of simple harmonic motions in two dimensions of a massless and chargeless particle in the static spherically symmetric (SSS) background. After solving the equations of motion numerically and analyzing the Poincar$\acute{e}$ sections we confirmed that the composite system shows chaotic trajectories when it comes under the influence of the event horizon of a SSS blackhole. Here we discuss about the PSD diagrams for this system to get an intuitive idea about the route to chaos.

 Now the total energy of the composite system under the background of static spherically symmetric spacetimes is \cite{Dalui:2018qqv}
\begin{eqnarray}
E=&&-\sqrt{1-f(r)}~p_r+\sqrt{p_r^2+\frac{p_\theta^2}{r^2}}+\frac{1}{2}K_r(r-r_c)^2+\frac{1}{2}K_\theta(y-y_c)^2~.
\label{App6}
\end{eqnarray}
and correspondingly, the equations of motion will have the form as 
\begin{eqnarray}
&&\dot{r}=\frac{\partial E}{\partial p_r} = -\sqrt{1-f(r)}+\frac{p_r}{\sqrt{p_r^2+\frac{p_\theta^2}{r^2}}}~;
\label{App7}
\\
&&\dot{p_r} = -\frac{\partial E}{\partial r} = -\frac{f'(r)}{2\sqrt{1-f(r)}}p_r+\frac{p_\theta^2/r^3}{\sqrt{p_r^2+p_\theta^2/r^2}}- K_r(r-r_c)~;
\label{App8}
\\
&&\dot{\theta} = \frac{\partial E}{\partial p_\theta} =\frac{p_\theta/r^2}{\sqrt{p_r^2+p_\theta^2/r^2}}~;
\label{App9}
\\
&&\dot{p_\theta} = -\frac{\partial E}{\partial\theta}=-K_\theta r_H(y-y_c)~,
\label{App10}
\end{eqnarray}                                  
where $f(r)$ is taken to be near horizon form $f(r) \simeq 2\kappa(r-r_H)$ with $\kappa$ and $r_H$ are surface gravity and location of horizon, respectively. In the above $y$ is chosen to be $y=r_H\theta$. Since here $r$ and $\theta$ are not Cartesian coordinates, the system can be called as a harmonic potential distorted to align it along a $r=$ constant circle.
\begin{figure}[!ht]
	\centering
	\includegraphics[scale=0.21, angle=0]{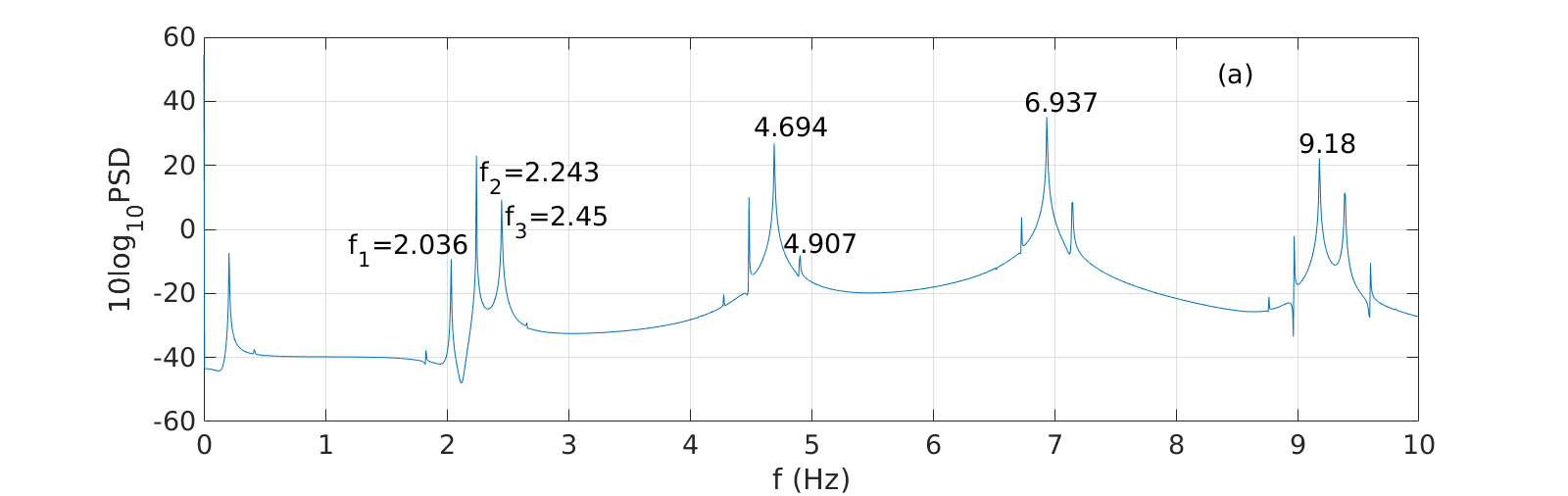}
	\includegraphics[scale=0.21, angle=0]{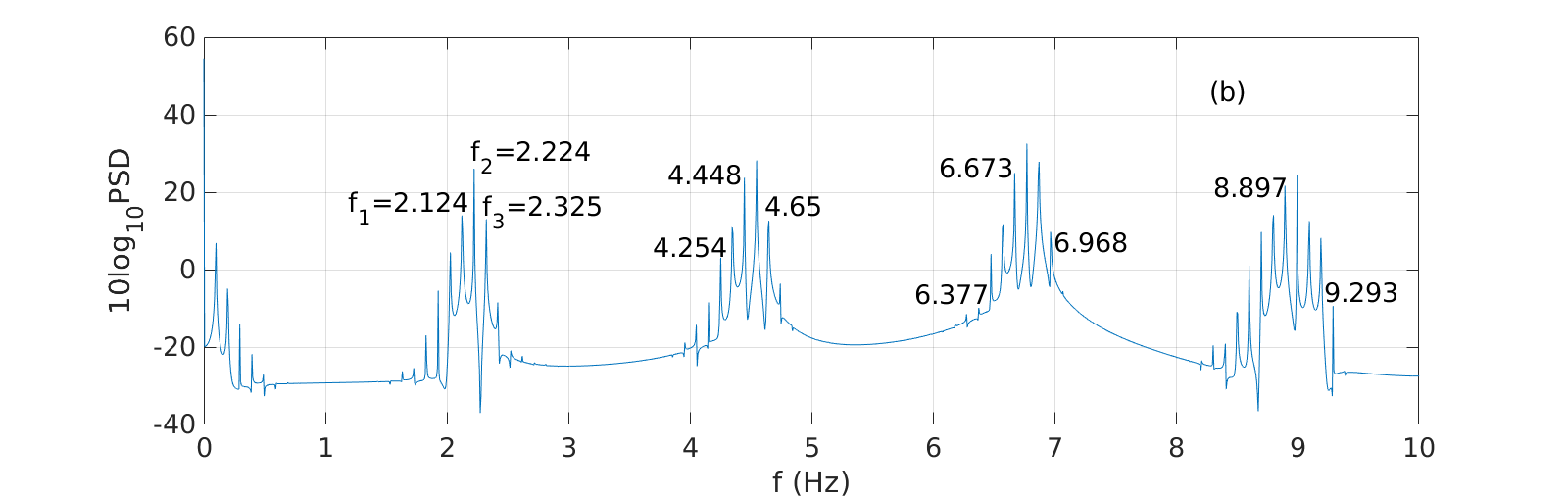}
	\includegraphics[scale=0.21, angle=0]{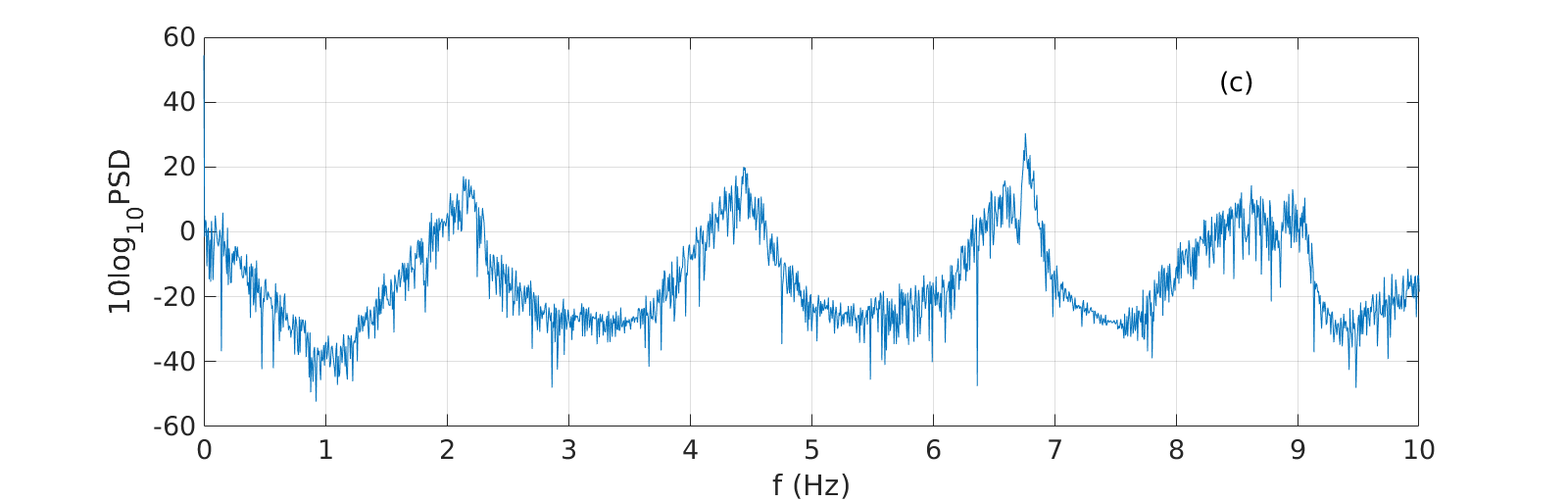}
	\includegraphics[scale=0.21, angle=0]{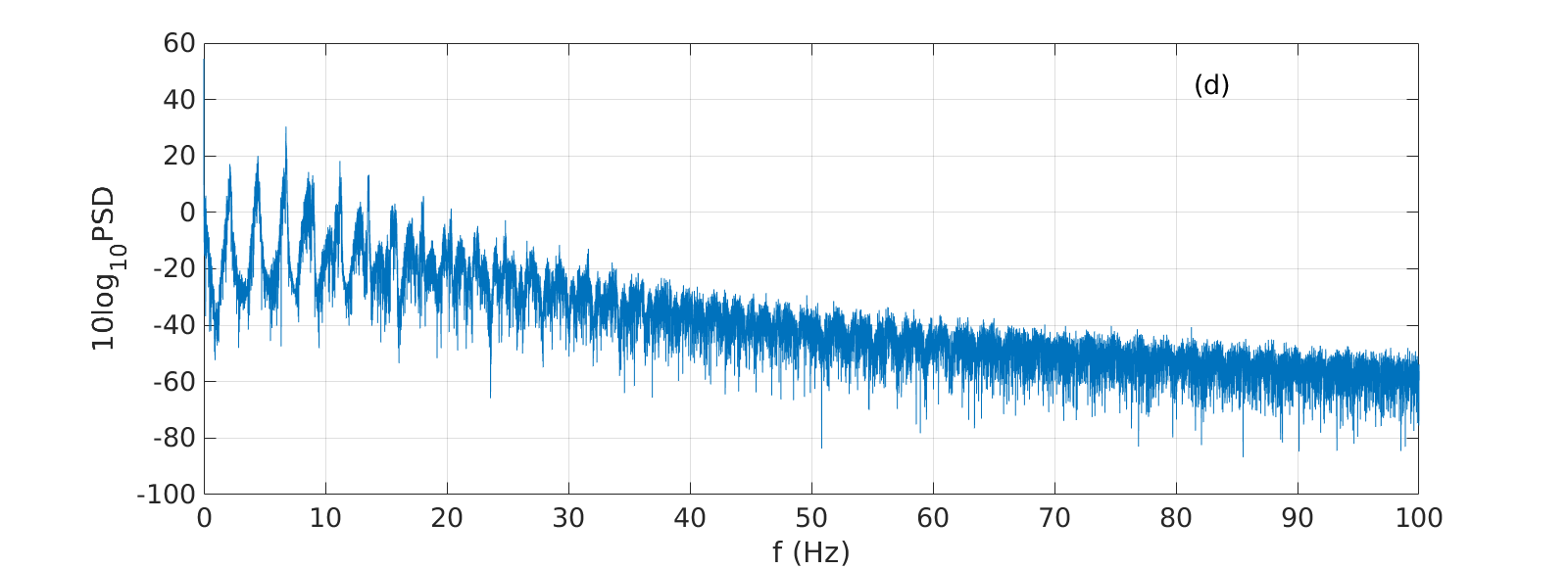}
	\caption{(Color online)  PSD for the SSS black hole at different values of energies with $\theta=0,\, p_{\theta}>0,\,K_{r}=100,\, K_{\theta}=25,\,r_{c}=3.2,\,\theta_{c}=0$, \textcolor{blue}{$\kappa=0.25$, $r_H=2.0$}. The enrgies are $(a)\,E=75,\,(b)\,E=77,\,(c)\,E=78.5$. Upto $E=75$, only $f_{1},f_{2}$ and $f_{3}$ and their harmonics are present but for higher values of energy i.e from $E=77$ onwards more frequencies start populating the spectrum. At $E=78.5$, the highly population of frequency spectrum and the exponential decay indicate $(d)$ the onset of chaos. }
	\label{fig:SSSPSD}
\end{figure}

After solving these equations (\ref{App7}-\ref{App10}) numerically we plot the PSD diagrams using (\ref{1.14}) for different values of the total energy of the system. The values of the other parameters are set as \cite{Dalui:2018qqv}. Now in Fig. \ref{fig:SSSPSD} we see for low energy value of the total energy of the system ($E=75$) the system contains only three frequencies and their harmonics (see Fig. \ref{fig:SSSPSD}$(a)$). As the total energy of the system gets increased the appearance of more frequencies are noticed (Fig.\ref{fig:SSSPSD}$(b)$) and at high value of $E=78.5$ (Fig. \ref{fig:SSSPSD}($c$)) the highly population of frequency spectrum indicates that the system is in the chaotic regime. 

In case of the accelerated frame we observed that for lower values of energy the PSD diagrams contain only three frequencies and their harmonics (see Fig. \ref{fig:PSD_a0p35}) which is similar to the present case (Fig. \ref{fig:SSSPSD}) where the particle is moving near the event horizon of a SSS black hole. So this observation implies that the route to chaos for both the cases are similar.  

\end{widetext}



\end{document}